\documentclass[twocolumn]{aastex631}

\shorttitle{Nova FQ Circini Has a B1 V(n)(e) Companion}
\shortauthors{Schaefer}
\graphicspath{{./}{figures/}}

\begin{document}
\title{FQ Circini: An Ordinary Nova with a High-mass B1 V(n)(e) Companion Whose Decretion Disk Transfers Mass to the White Dwarf via Roche-Lobe Overflow.}

\author[0000-0002-2659-8763]{Bradley E. Schaefer}
\affiliation{Department of Physics and Astronomy, Louisiana State University,
Baton Rouge, LA 70803, USA}

\author{Andrew Pearce}
\affiliation{American Association of Variable Star Observers,
35 Viewway, Nedlands, Western Australia 6009, Australia}

\author[0009-0000-7419-8118]{Tom Love}
\affiliation{Variable Stars South, Royal Astronomical Society of New Zealand,
PO Box 3181, Wellington 6011, New Zealand}
\affiliation{Centre for Astrophysics, University of Southern Queensland, 
Toowoomba, QLD 4350, Australia }

\author[0000-0003-0155-2539]{Michael M. Shara}
\affiliation{Department of Astrophysics, American Museum of Natural History, 
New York, NY 10024, USA}

\author[0000-0001-9788-3345]{Lee Townsend}
\affiliation{South African Astronomical Observatory, 
Observatory Road, Observatory 7925, Capetown, South Africa}
\affiliation{Southern African Large Telescope, 
Observatory Road, Observatory 7925, Cape Town, South Africa}

\author[0000-0002-5648-3107]{Simon J. Murphy}
\affiliation{Centre for Astrophysics, University of Southern Queensland, 
Toowoomba, QLD 4350, Australia }

\author[0000-0001-6797-887X]{Christopher J. Corbally}
\affiliation{Vatican Observatory Research Group, 
University of Arizona, Tucson, AZ 85721-0065, USA}

\begin{abstract}

FQ Cir was an ordinary fast He/N classical nova, peaking at $V$=10.9.  The pre-eruption and post-eruption counterpart was at $V$=14.0, making the smallest known classical nova amplitude of 3.1 mag.  The nova light and the counterpart coincide to 0.034 arc-seconds, and the counterpart is a rare hot/blue emission-line star with flickering, so the identification of the quiescent nova is certain.  The counterpart is a weak Be main sequence star, B1 V(n)(e).  A coherent photometric period appears in all four {\it TESS} Sectors and in the AAVSO post-eruption light curve, as ellipsoidal modulation with orbital period 2.041738 days.  The companion must have been spun-up to a fast rotation, and like all Be stars, a decretion disk is exuded.  With the constraints of the blackbody radius and the main sequence, the companion mass is 13.0$^{+0.2}_{-0.5}$ $M_{\odot}$, with radius 6.2$\pm$0.2 $R_{\odot}$.  This is the discovery of a cataclysmic variable with a high-mass companion, a new class that we call `High Mass Cataclysmic Variables'.  The white dwarf mass is 1.25$\pm$0.10 $M_{\odot}$ and must have an accretion disk that supplies fuel for the nova eruption.  FQ Cir represents a new mode of accretion in interacting binaries, with Roche lobe overflow from the decretion disk feeding mass into the usual accretion disk  around the white dwarf, for disk-to-disk accretion.  From the mass budget of the binary, the primary star must have its initial mass $>$7.7 $M_{\odot}$, forming an ONe white dwarf, so FQ Cir can never become a Type Ia supernova.

\end{abstract}

\section{INTRODUCTION}

All cataclysmic variables (CVs) are close interacting binaries where a usually ordinary low-mass star ($\lesssim$1 $M_{\odot}$), called the `companion star' or the `donor star', feeds gas through an accretion disk onto a white dwarf (WD) by the mechanism of Roche lobe overflow. (RLOF)  All CVs have the gas accumulated on the WD surface eventually reaching the trigger condition for a thermonuclear runaway explosion (Shara 1981) that appears as a classical nova (CN, Chomiuk et al. 2020).  CNe that have more than one observed nova eruption are labeled as recurrent novae (RNe).  Dwarf novae and nova-like systems are those CVs for which the last eruption was long ago, or perhaps missed in the last century. 

Many types of CV companion stars have been recognized (Schaefer 2022a).  The most common type of companions are low mass main sequence stars, with spectral classes K0--M9.  To fill their Roche lobes, these binaries must have short orbital periods ($P$).  For these stars, the observed orbital period gives a measure of the companion mass.  The hottest and most massive companion stars on the main sequence are something like G0 at $\lesssim$1 $M_{\odot}$, with orbital periods of around 0.6 days.  Longer orbital periods are seen, with 18\% of all novae having periods 0.6--10 days, where the companion is a subgiant, always with a surface temperature under 6000 K or so, and with a mass around 1 $M_{\odot}$ or so.  Finally, 13\% of all novae have orbital periods from 10--1000 days with ordinary red giant companions, all with low surface temperatures and with masses $\lesssim$1 $M_{\odot}$.  With one exception\footnote{The one exception is the weird and extreme V Sge (Schaefer, Frank, Chatzopoulos 2020), with a companion surface temperature of roughly 12,000 K.  This binary is in a 0.514 day orbit of a 0.85$\pm$0.05 $M_{\odot}$ WD with RLOF from a 3.9$\pm$0.3 $M_{\odot}$ companion.  The uniquely-high mass ratio and the RLOF means that the binary is even now in the middle of exponential runaway accretion, with a doubling timescale of 89 years.  V Sge is the only known case of what we call an `intermediate mass cataclysmic variable' or IMCV.}, all CV companion stars have surface temperatures $\lesssim$6000 K and masses $\lesssim$1 $M_{\odot}$.

This stark situation for CVs can be contrasted to the case for close interacting binaries with a neutron star (NS), or black hole (BH) in place of the WD.  These systems are called X-ray binaries (XRBs).  These systems are neatly trimodal on the mass of the companion star.  The low-mass XRBs (the LMXBs) all have $\lesssim$1 $M_{\odot}$ companions, with prototypes Sco X-1 and A0620-00.  The  high-mass XRBs (the HMXBs) all have $>$10 $M_{\odot}$ companions, with prototypes Cyg X-1 and Vela X-1.  There are rare intermediate-mass XRBs (the IMXBs) with $\sim$3 $M_{\odot}$ companions, with prototypes Her X-1 and V4641 Sgr.

Where are all the CVs with high mass companions?  The obvious group name is `high-mass cataclysmic variables'', or `HMCVs'.  That is, where are all the close interacting binaries that consist of WDs accreting from $>$10 $M_{\odot}$ companions?  HMCVs should be more luminous than the common CVs, so they will have a higher probability of recognition and they will be visible over a larger volume of our Milky Way, so they are not being `hidden'.  As the gas accumulates onto the surface of the HMCV WD, it will inevitably trigger an ordinary CN eruption, with this eruption necessarily appearing with a remarkably low eruption amplitude.

In this paper, we report the discovery of an HMCV, which has exploded as an ordinary nova in 2022, named FQ Cir.  The WD has a mass around 1.25 $M_{\odot}$ being fed by an ordinary CV accretion disk, so as to provide the trigger mass for the nova.  But the companion star is a Be main sequence star with a mass near 13.0 $M_{\odot}$.  Startlingly, this makes FQ Cir into an HMCV.  The orbital period is 2.04 days, so the Be star must be spun-up and ejecting the decretion disk that is truncated by the Roche lobe, such that gas in the decretion disk will pass through the Roche lobe and fuel the WD's accretion disk.  So we have a new configuration for accretion, with two separate disks in the system, with one disk (the decretion disk around the B-star) transferring gas to the second disk (the accretion disk around the WD) by RLOF.

After this paper was largely written, while awaiting only the last spectra from Siding Spring, Chamoli et al. (2025) published an excellent paper with a similar conclusion.  Their target was a rapid recurrent nova in the Andromeda Galaxy.  They conclude ``Combined with archival spectra consistent with a B-type star with H$\alpha$ in emission, this points to the quiescent counterpart being a Be star with a circumstellar disk. We propose that M31N 2017-01e arises from a {\it rare BeWD binary}, where the white dwarf (WD) accretes from the decretion disk of its companion, explaining its rapid recurrence, low-amplitude outbursts, and unusual quiescent luminosity and color.''

In Section 2 of this paper, we measure and derive all the properties of the FQ Cir binary.  Subsequent sections analyze a variety of big-picture questions.  Section 3 shows FQ Cir to be an HMCV and to have the new disk-to-disk accretion mode.  Section 4 demonstrates that the initial mass of the primary star was $>$7.6 $M_{\odot}$, which necessarily produces a CO WD, and it cannot produce a Type Ia supernova.  Section 5 sketches out the evolution of the FQ Cir binary, with a specific and detailed model being needed.  Section 6 gives a summary of the nature of the FQ Cir binary.  Section 7 lists seven specific tasks for future studies.

\section{NOVA FQ CIR PROPERTIES}

This paper is about the ordinary classical nova FQ Circini (FQ Cir, Nova Cir 2022, TCP J15244460-6059200) that has an observed amplitude of 3.1 mag.  This section reports our observations and collects archived observations, with these measuring the properties of FQ Cir.

\subsection{Nova Eruption}

FQ Cir was discovered by A. Pearce (Nedlands, Western Australia) on 2022 June 25.48 at unfiltered magnitude of 10.7.  (It was not visible on his images taken 1.0 days earlier.  Nor was the nova visible 2.5 hours before discovery, to a limit 3 mags below the discovery magnitude, as observed by R. McNaught, Coonabarabran, NSW Australia.)  The peak was at magnitude 10.9, and it faded by two magnitudes in the $V$-band in close to 2.0 days ($t_2$=2 days), and faded by 3.0 mag in 11 days.  The observed light curve is apparently the sum of the nova light plus the light from a 14th-mag star, so the nova light has a decline rate of $t_3$$\le$11 days.  The light curve is smooth, so the light curve class is S(11), with the 11 being an upper limit.  

Love (2022) and Aydi et al. (2022) report that the spectrum at peak ``shows a red continuum with broad and strong emission lines of H I, He I, and OI'', with no P-Cygni line profiles, and with the H$\alpha$ FWZI covering 10000--14600 km/s.  The Aydi spectrum is not available, but our 4 eruption spectra (26 June to 3 July 2022) are visible and downloadable at the ARAS data archive\footnote{\url{https://aras-database.github.io/database/novacir2022.html}}.  Love (2022) classified FQ Cir as a fast He/N nova, as based on the helium and oxygen lines.  This eruption spectroscopy is important for proving that the 2022 event is a classical nova and hence that the FQ Cir binary must have a WD star accreting from a companion.

The spectrum at peak has a red continuum with broad Balmer lines and He I, while our spectra  also show O I emission lines (Love 2022).  These spectra show H$\alpha$ with a full-width-at-half-maximum (FWHM) of 2600 km s$^{-1}$.  So FQ Cir is a normal CN of the He/N spectral classification.  This conclusion is important as it is proof that an accreting WD exists in the counterpart.

The nova light faded extremely fast, and after 12 days, the brightness reached an asymptote to a flat light curve with $V$=14.0.  What was left visible was a constant star with the colors, spectrum, and luminosity of an ordinary star with spectral class B, which we will name as the `14th-mag' star or as the `B-star'.  This same star was quickly recognized in many pre-eruption data sets.  So formally, the nova amplitude is $14.0-10.9$, or 3.1 mag.  This is startling, as this would be by far the smallest amplitude classical nova eruption\footnote{Schaefer (2025b) report all known nova amplitudes for the 402 Milky Way novae, and the smallest amplitude is 5.1 mag for the utterly unique helium nova V445 Pup.  For the smallest amplitude of all the rest, we have 5.6 mag for V1534 Sco, for which the small amplitude is due to its red giant companion star.  The next two smallest amplitudes are for V794 Oph and RS Oph, both of which have red giant companions.  The smallest amplitude for any nova without a known red giant companion is 6.9 mag for V1655 Sco.}.  This opens the central mystery and excitement for FQ Cir.

\subsection{Light Curve Trends, Flares, and Flickering}

\begin{figure*}
	\includegraphics[width=2.1\columnwidth]{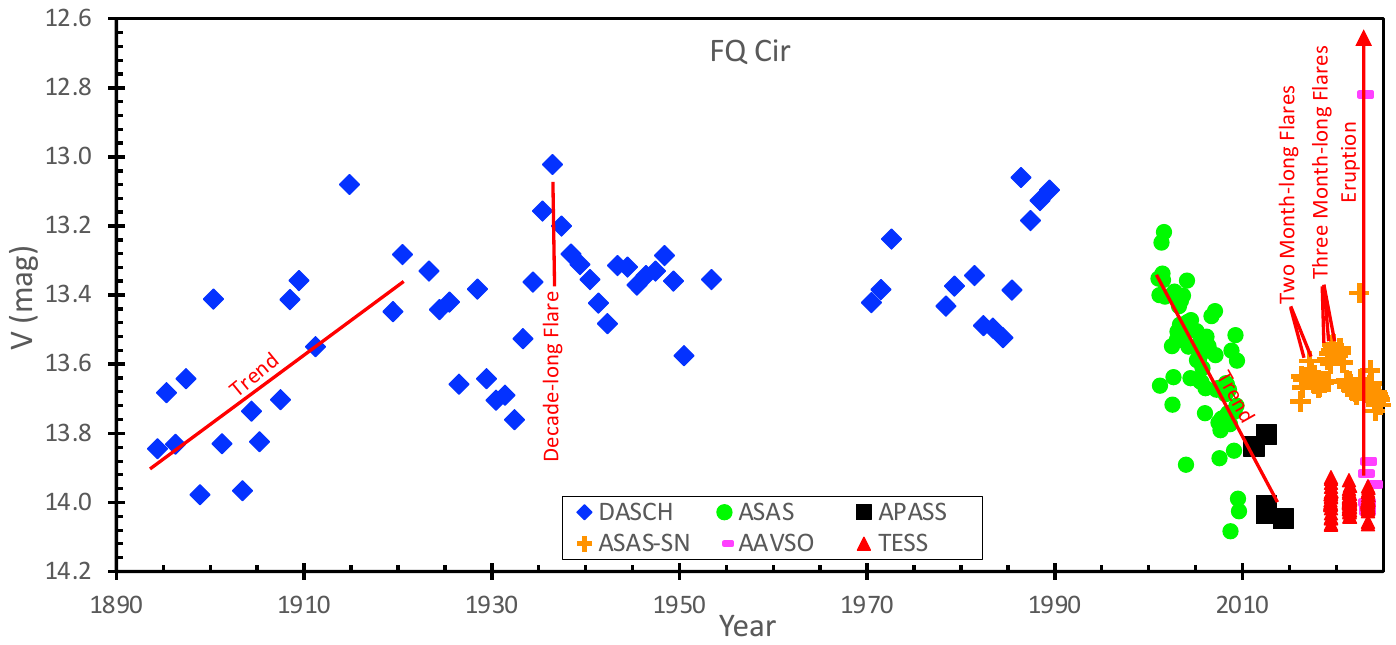}
    \caption{Light curve for FQ Cir 1894--2025.  The $V$ magnitudes represent the brightness only of the B-star, because the WD and disks are greatly fainter than the hot star.  This light curve includes magnitudes from DASCH 1-year averages (blue diamonds), ASAS 0.1-year averages (green circles), APASS magnitudes (black squares), ASAS-SN 0.05-year averages (orange plus signs), AAVSO 0.05-year averages (magenta dashes), and {\it TESS} 1-day averages (red triangles).  The two red lines show the two most prominent trends that last longer than one decade.  The six labeled flares show the cases of well-resolved and significant flares, variously month-long and decade-long.  The 2022 nova eruption is indicated with the red up-arrow.  One of the points from this light curve is that the 14th-mag star has long-term trends and flickering on all time scales, with this being impossible for all {\it normal} B-stars, yet such is seen in all Be stars.  Such a rare star is incredibly unlikely if the 14th-mag star is a random foreground star.  Another point is that the star is not systematically brightening at an accelerating rate, so this indicates that the high-mass-ratio binary is not accreting by ordinary Roche lobe overflow from a companion filling its Roche lobe (because such would require runaway accretion).  }
\end{figure*}

\begin{figure*}
	\includegraphics[width=2.1\columnwidth]{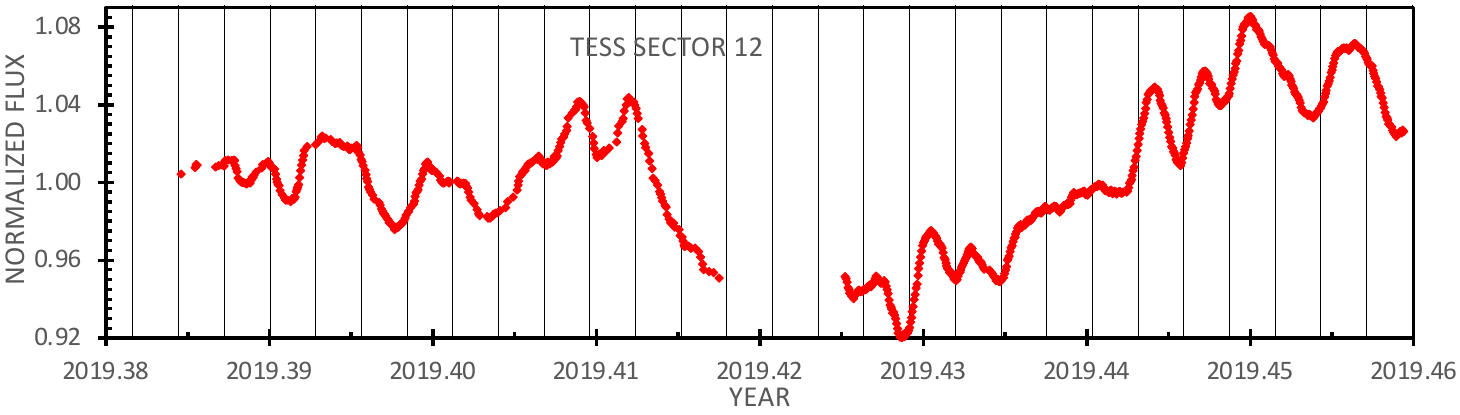}
	\includegraphics[width=2.04\columnwidth]{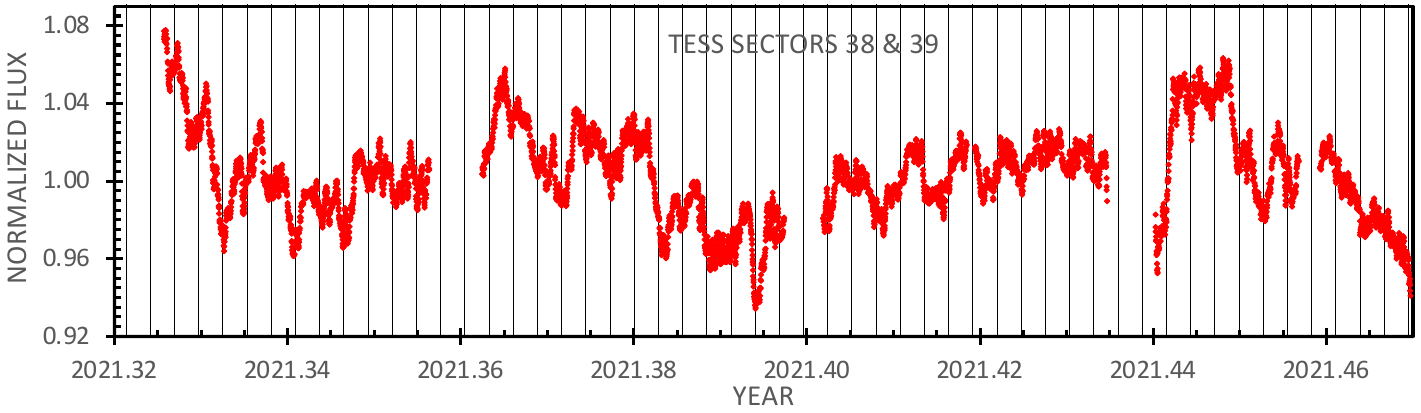}
	\includegraphics[width=2.1\columnwidth]{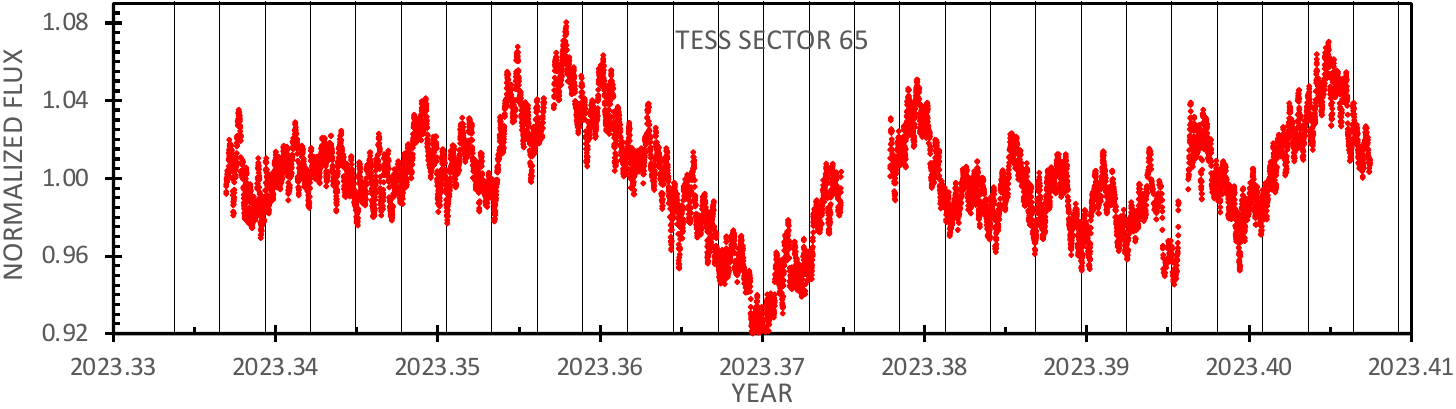}
    \caption{{\it TESS} light curve for FQ Cir.  These three panels plot the {\it TESS} normalized flux for Sector 12 (top panel), Sectors 38 and 39 (middle panel), and Sector 65 (bottom panel).  These are smoothed with a simple running average.  The dominant feature is that of $\sim$5\% flickering on timescales of hours, days, and weeks.  Such behavior is never seen for {\it normal} B-stars with no emission lines.  Such behavior is always seen for Be stars.  That a unique hot-blue star happens to be within 0.034 arc-seconds of the nova position is highly improbable -- unless the star is the nova counterpart.  Further, the {\it TESS} light curves show a highly significant photometric periodicity (1.020869$\pm$0.000018 days) superposed on the flickering.  The thin vertical lines indicate the times of minimum light according to the best fit ephemeris, with epoch BJD 2459740.9710$\pm$0.0065.  In general, the times of local minima are close to the lines, while the local maxima are between the vertical lines, matching the ephemeris.  Nevertheless, often there is little apparent connection with the ephemeris.  The periodicity is highly significant, see the Fourier transform in Figure 3, so the ubiquitous flickering is often hiding the sinewave modulation.  The 1.02 day photometric modulation is caused by ordinary ellipsoidal modulation, for which the orbital period is twice the photometric period, so $P$=2.04 days.  See the average folded light curve in Figure 4.  This periodicity is coherent and stable over 6.3 years, and the only such clock is the orbit.   }
\end{figure*}

The 14th-mag B-star at the nova position has many pre-eruption and post-eruption magnitudes, so we construct light curves with fairly high density from 1894 to 2025.  This is plotted in Figure 1.

The Harvard College Observatory collection of archival sky photographs (plates) recorded FQ Cir from 1894--1989.  The plates show the field, where the size of the target image is compared to the sizes of nearby comparison stars of known magnitude.  The plates have a spectral response identical to the modern $B$-band system, and the comparison stars are exactly in the modern $B$-band system, so the measured magnitudes are exactly in the modern $B$-band system.  The magnitudes for individual plates were extracted by the Digital Access to a Sky Century @ Harvard (DASCH) program\footnote{\url{https://dasch.cfa.harvard.edu/}}, lead by J. Grindlay (Harvard).  The real uncertainty is each measure is typically $\pm$0.15 mag.  Importantly, the listed magnitudes must have stringent quality control.  In particular, plates with red- and yellow-sensitivity must be rejected, plates with $>$0.30 mag uncertainty and within 0.30 mag of the plate limit are rejected, and the various types of plate defects are best rejected by requiring the image centroid to be close to the target position.  The result is 656 $B$ magnitudes 1894--1989.  For plotting purposes in Figure 1, these $B$ magnitudes are converted to $V$ by subtracting the average $B-V$ of $+$0.46 mag.  For plotting purposes, we have averaged the magnitudes into 1.0-year time bins, with these shown in Figure 1 as blue diamonds.  After 1925, the error bars on the yearly averages are typically $\pm$0.04 mag.  This is the distance between the minor tick marks in Figure 1, so the flares and variability are significant.

The All Sky Automated Survey\footnote{\url{https://www.astrouw.edu.pl/asas/}} (ASAS) lead by G. Pojmanski (Warsaw University) covered the entire sky down to $V$=15 from 2001--2009.  For FQ Cir, ASAS has 618 $V$-band magnitudes of Grade A or B, for which we use the aperture for `MAG\_2' (as having the smallest quoted uncertainty).  The quoted photometric error bars are typically $\pm$0.04 mag.  The RMS scatter around any smooth trend is 0.15 mag, which might be from unresolved flickering in this sparse light curve.  Figure 1 plots the ASAS light curve, with 0.10-year time binning, as green circles.  Strikingly, the light curve shows a secular fading trend that averages 0.5 mag over 8.5 years.

The All-Sky Automated Survey for Supernovae (ASAS-SN; Hart et al. 2023) covers the entire sky to $V$$\sim$17.5 with thousands of $V$ and $g$ magnitudes from 2016 to the present.  The FQ Cir light curve\footnote{\url{http://asas-sn.ifa.hawaii.edu/skypatrol/objects/652835699853}} has 200 $V$ magnitudes from 2016--2018 and 889 $g$ magnitudes from 2018--2025 (with this including 10 points in the 2020 eruption).  The quoted measurement errors are typically 0.007 mag.  For plotting purposes, we have converted the $g$ measures into $V$ with $\langle g \rangle$-$\langle V \rangle$=0.23, as taken from their year of overlapping data.  The ASAS-SN light curve, averaged over 0.1-year time bins, is plotted as orange plus-signs in Figure 1.  The cadence and coverage of the ASAS-SN light curve are good for measuring well-resolved month-long flares, with the most blatant being two in 2017 and three in 2019.  These flares have amplitudes of 0.1--0.2 mag.  In Figure 1, we see that the ASAS-SN light curve is offset from the simultaneous AAVSO data as being brighter by 0.3 mag.  This same offset is seen versus other magnitudes in quiescence, for example with SkyMapper pre-eruption magnitudes (Aydi et al. 2022).  Further, various official versions of the ASAS-SN light curve are inconsistent with each other at the 0.3 mag level, and we are likely seeing the ordinary problems of correcting for the color term of a very blue star.  So we think that the plotted ASAS-SN light curve really should be systematically offset fainter by 0.3 mag or so.  This uncertainty does not affect any of the science or conclusions.

The American Association of Variable Star Observers (AAVSO) has run an all-sky survey covering $BVgri$ magnitudes for almost all stars down to around 19th mag, with the resultant catalog being named the AAVSO Photometric All-Sky Survey\footnote{\url{https://www.aavso.org/apass}} (APASS), lead by A. Henden (AAVSO).  These magnitudes are definitively calibrated against the Landolt standard stars.  APASS is unique and valuable for a variety of purposes, including providing accurate comparison star magnitudes for FQ Cir, DASCH, and all novae, as well as providing well-measured magnitudes for FQ Cir.  For FQ Cir, the catalog gives $BVgri$ magnitudes for 6--8 nights.  With the $B$ and $V$ measures taken simultaneously, the $\langle B-V \rangle$=0.47$\pm$0.05 measures are good for avoiding the jitter arising from flickering.  The 6 $V$ measures are plotted in Figure 1 as black squares.

The AAVSO has collected 54 million magnitude measures from around the world for long-term light curves of most named variable stars, all included in the AAVSO International Database (AID), with easy public access\footnote{\url{https://www.aavso.org/data-access}} for all the light curves.  For FQ Cir in particular, the AAVSO has the most complete eruption light curve, with 120 $BVRI$ magnitudes in the first week of the eruption.  After the eruption, the light curve has 1803 $V$ magnitudes in quiescence.  Figure 1 has the AAVSO $V$ light curve (binned to 0.05-year averages) plotted as magenta dashes.

The {\it TESS} satellite (led by G. Ricker, MIT) has been producing wonderful light curves of almost all stars in the sky, down to 19th mag, with these having time series with many Sectors of data.  Each Sector lasts nearly 27 days of continuous photometry with few gaps, with time resolution often as good as 200 seconds.  For FQ Cir, {\it TESS} observed four Sectors; 12 (early 2019), 38, 39 (2021), and 65 (post-eruption in 2023).  These data are publicly available at the Barbara A. Mikulski Archive for Space Telescopes\footnote{\url{https://mast.stsci.edu/portal/Mashup/Clients/Mast/Portal.html}} (MAST).  The photometry is reported as a flux (in units of counts per second) covering 21$\times$21 arc-second pixels, with a broadband sensitivity roughly 6000--10,000~\AA.  For FQ Cir, we have adopted the standard QLP light curve, optimized for correcting the various systematic problems.  The light curves for all four Sectors are plotted in Figure 2.  For example, Sector 65 records 10,206 fluxes with 200 s time resolution covering 25.8 days with few gaps (see the bottom panel of Figure 2).  These fluxes averaged over 1-day bins (converted to magnitudes with the average set to 14.0 mag) are plotted as red triangles in Figure 1.

The light curves 1894--2025 show that FQ Cir has variability on all time scales, with amplitude 0.05--0.5 mag.  The light curve shows a broad rise, flat-top, and decline lasting from 1894 to 2015.  The light curve displays roughly-linear trends, for example a rise by 0.5 mag from 1894 to 1920, and a decline by over 0.6 mag from 2001 to 2014.  The light curve displays discrete flares, with well-resolved examples including a 0.74 mag flare from 1932--1942 and two month-long flares in 2017 with amplitude 0.06 mag.  The {\it TESS} light curves show continuous flickering with timescales from one day to one week.  Looking at the fine detail, we see flickering down to the one hour timescale.  In all, FQ Cir shows factor-of-two variability on all timescales from one hour up to over one century.

This photometric history is unique amongst normal B stars.  If any normal B star had ever been seen to show any such variability, it would have been called out and recognized long ago.  Despite many thousands of B stars with adequate variability searches, we know of no precedent for a normal B star having century-long trends, year-long or month-long flares, or incessant flickering on timescales down to one hour.  To quantify this, we can take the sample of the 252 B0--B5 normal main sequence stars in the Yale Bright Star Catalogue (Hoffleit 1982, Hoffleit \& Saladyga 1997), with this sample being all the brightest and best observed B stars that have the best variability searches on all timescales.  This sample has 19 known variable stars, of which 13 are eclipsing binaries, and the remaining 10 variables are periodic small amplitude ($<$0.06 mag) variables.  So zero-out-of-252 have trends, flares, or flickering.  This measured limit shows that the observed variability of the B-star is unique at the $<$0.4\% fractional level.  The actual rarity of this phenomenon is greatly smaller than this 0.4\% limit, because all normal B stars brighter than 10th mag have also been adequately tested for trends/flares/flickering with no discovered cases.  This unique behavior of the B-star is very improbable in the case that the B-star is a random foreground field star unrelated to the nova.

While {\it normal} B stars have no such variability, a common class of B stars with prominent emission lines do show variability with trends and flares on all timescales.  These are the Be and B[e] stars, defined as B stars with emission lines (Porter \& Rivinius 2003).  The emission is primary visible in the Balmer lines, most prominent in H$\alpha$, and the emission varies greatly in strength, from huge emission lines down to mere minor filling-in of the Balmer lines.  These classes are heterogeneous, being composed of pre-main-sequence stars, supergiants, symbiotic stars, and planetary nebula nuclei, as well as main sequence stars.  The common property of these systems is that the Be star is surrounded by a dense disk or shell of gas that produces the emission lines.  For the classical Be stars on the main sequence, a characteristic mechanism is that the star is rotating so fast (sometimes near the breakup velocity) that it ejects a `decretion disk', typically extending out to 20 or 40 $R_{\odot}$ (Rivinius, Carciofi,  \& Martayan 2013).  If FQ Cir has a Be star companion, then we have an easy precedent and understanding for its variability.  The B-star in FQ Cir does not show blatant emission lines above the continuum, but it does show filling-in of the Balmer lines (see Section 2.10).  This emission only confirms the Be star nature of the companion, which we already know from the large-amplitude flares/trends/flickering.  So FQ Cir is a weak Be star, and has a decretion disk.

The 132-year light curve shows that the brightness of FQ Cir is {\it not} increasing exponentially.  Such a case is expected if the counterpart has accretion by RLOF.  That is, with Kepler's Law, a binary with a mass ratio $q$ ($M_{\rm comp}$/$M_{\rm WD}$) that is $\gg$1 must have an exponentially increasing accretion rate (Frank, King, \& Raine 2002, Temmink et al. 2023).  Such a runaway accretion and exponential brightness increase (with an exponential timescale of 89 years) has been observed for only one binary, V Sge with $q$=3.9, and this is the only known RLOF binary with $q$$\gg$1 (Schaefer, Frank, Chatzopoulos 2020).  Figure 1 shows that FQ Cir is not brightening exponentially on a time scale of a century or more, so FQ Cir must not be having RLOF from the companion surface touching the Roche lobe.  Yet still the WD needs accretion so as to feed the nova event, and a simple way to provide the gas is from the decretion disk.  The gas in the decretion disk is necessarily spiraling out and must inevitably cross the companion's Roche lobe, then fall onto the accretion disk around the WD, ultimately falling onto the WD so as to accumulate and power the next nova eruption.

\subsection{Photometric Periodicity}

\begin{figure}
	\includegraphics[width=1.01\columnwidth]{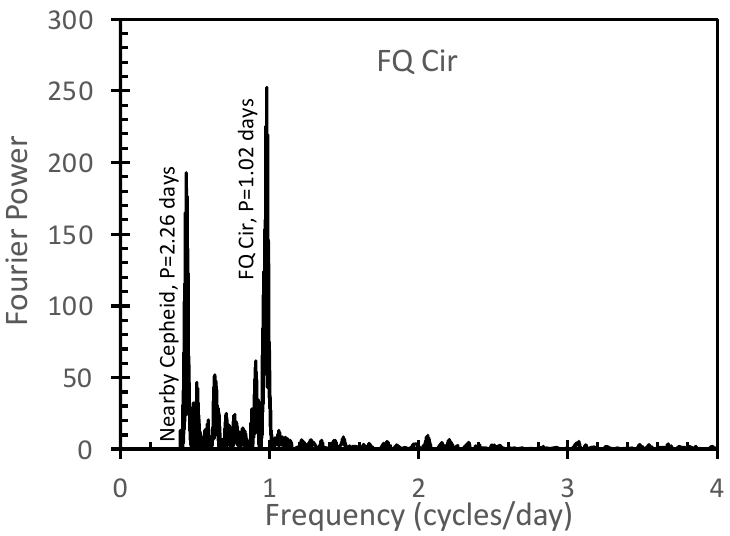}
    \caption{Fourier transform of the {\it TESS} light curve.  The primary point of this figure is that a highly significant Fourier peak (at a frequency of 0.98 cycles/day, with photometric period 1.02 days) appears centered on the nova pixel.  This figure is shown so that it can be seen that the Fourier peak is isolated and high above the local background noise level, demonstrating that the peak is highly significant.  This high significance is also proven by the chi-square fit to the light curve (c.f. Figure 4).  This same peak is independently visible with the same strength and period in each of {\it TESS} sectors 12, 38, 39, and 65.  No significant Fourier peak is seen for the double-period of 2.04 days (for a frequency of 0.49 cycles/day).  The Fourier peak could arise from an orbital period of 1.02 days with irradiation and hotspot effects dominating over the ellipsoidal effect, or from an orbital period of 2.04 days with the ellipsoidal modulation dominating over any irradiation and hotspot effects.  With the hot B-star swamping the irradiation and hotspot effects, the ellipsoidal mechanism must be dominating, so the orbital period is actually $P$=2.04 days.  This plot also shows another significant peak for a period of 2.26 days, also centered on the nova pixel, with the folded light curve showing the characteristic shape of a Cepheid, so this is just a random foreground variable star near to the nova.  For frequencies lower than 0.4 cycles/day, the background noise starts rising due to the chaotic flickering on longer time scales.  }
\end{figure}

\begin{figure}
	\includegraphics[width=1.01\columnwidth]{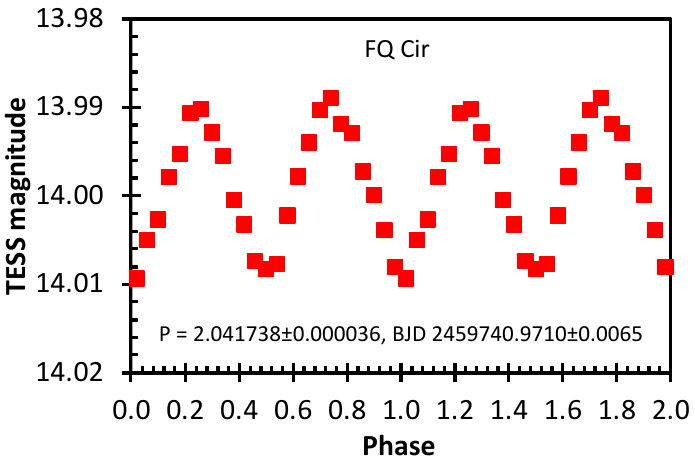}
    \caption{Folded light curve for the orbital periodicity.  Here, the {\it TESS} magnitude is the flux converted to magnitude with the average flux corresponding to 14.00 mag.  The light curves for individual cycles have their baselines varying from 13.92 to 14.08 because the counterpart is flickering up and down.  But when the 50 cycles are averaged together, a blatant sinewave periodicity is seen.  The photometric period is 1.020869 days for a simple sinewave, but the orbital period is twice that at $P$=2.041738 days for a double sinewave.  For FQ Cir, the ellipsoidal mechanism makes for two identical photometric minima each orbit, at the times of the inferior and superior conjunctions.  This figure plots the phase-averaged folded light curves in bins of 0.04, with the individual points having error bars of $\pm$0.002 mag.  The phase is calculated with the orbital period of 2.041738 days with a zero-phase for an epoch of minimum light of BJD 2459740.9710.  Each point is plotted twice, once with phase 0.00--1.00, and a second time with a phase 1.00 larger.  }
\end{figure}

Schaefer (2022a, 2025b) has been presenting results of a systematic search for previously undiscovered orbital periods in all Galactic CNe.  The basis of this program is that all novae and CVs show coherent and stable periodic modulations on the orbital period.  (The only exception is for low-inclination binaries.)  These modulations are always something close to sinusoidal light curves, or sinewaves with superposed eclipses.  The method is the simple construction of Fourier transforms, and looking for a significant isolated Fourier peak that appears identically in two-or-more independent data sets.  If such a peak is found, a full formal chi-square fit is made to the light curve, and this can serve as a test for significance, stability, and coherence.  The primary light curve source has been {\it TESS} photometry.  So far, {\it TESS} data have allowed the discovery of 35 new nova orbital periods.

As part of this program, we have recently searched the four Sectors of {\it TESS} light curves for FQ Cir.  The FQ Cir Fourier transform is shown in Figure 3.  This shows {\it two} isolated and highly-significant peaks, at periods of 1.02 days and 2.26 days.  The 2.26 day periodicity has a folded light curve that displays the characteristic shape of a Cepheid variable (i.e., a sinewave with a distinct fast rise and slow decline), so this can only be from one of the several stars inside the 21''$\times$21'' pixel, so this is a foreground star with no connection to the nova.  The 1.02 day photometric periodicity is the same in amplitude and period throughout each of the four Sectors.  The point of showing this Fourier transform is to provide an obvious visual confirmation of the significance of the periodicity.  This might be needed because the raw light curves have substantial flickering on all timescales that can often hide the orbital minima, or can create maxima that are not connected to the orbital ephemeris.  This is just a long-standing problem for many CVs with orbital modulation superposed on larger-amplitude flickering, where the by-eye looking at the light curve can make the modulation not-blatant.  The Fourier transform effectively ignores the flickering at low-frequencies, so the consistent and persistent orbital modulation becomes obvious in the plotted Fourier power.

We have made a chi-square fit of the light curve to a model sinusoidal light curve.   The center of the sinewave is set to the average flux, and the start epoch is set to a time that is the average of all the input times.  The free parameters are the period and the epoch of minimum light, with these varied until the chi-square is minimized.  It works out that the best fit has a reduced chi-square satisfactorily near to unity.  The one-sigma range for the period and epoch are calculated as the range over which the chi-square is within 1.0 of the minimum.  The fit for all four {\it TESS} Sectors (12, 38, 39, \& 65) gives a period of 1.020869$\pm$0.000018 days, and an epoch of minimum of BJD 2459740.9710$\pm$0.0065.  The folded light curve shape is reasonably close to a sinewave.  The full amplitude is 0.02 mag, but this is a a lower limit because the {\it TESS} pixels contain other (much fainter) stars plus background light.  For a double-period, the odd and even minima have identical depths, with this confirmed by the lack of any Fourier peak at 2.04 days (see Figure 3).  The chi-square for the zero-amplitude case is 557.5 larger than the best fit case (with the number of degrees of freedom equalling 17595), so the formal significance of the periodicity is 24-sigma.

This same periodicity is significant in the AAVSO post-eruption data, with 1757 $V$ magnitudes from 2023--2025, taken by the veteran observer F.-J. Hambsch (Oude Bleken, Belgium).  In this case, with the magnitudes all measured from one observatory, the $P$=1.02 day signal has its wings interfered by the 1-day window function.  So this is not a clean detection of the periodicity, yet it can provide an independent confirmation of the strong {\it TESS} periodicity.  This can also provide the connection from the periodicity to the 14th-mag B-star (rather than to some nearby foreground star).  Further, the photometry is of the B-star alone, rather than of the 21''$\times$21'' {\it TESS} pixel filled with unknown background flux, so the AAVSO data provide a good measure of the full amplitude of variability, which is seen to be near 0.028 mag in the $V$-band.

The 1.02 day periodicity is stable and coherent across 6.3 years (2019--2025).  The only astrophysical clock that is so stable and coherent in that range of periods is the orbital period.  So the photometric period is 1.020869$\pm$0.000018 days, and this must be tied exactly to the orbital period.

\subsection{Orbital Period}

The observed photometric periodicity of 1.02 days is certainly tied exactly to the orbit, but there are two orbital periods that can produce the observed data.  The first possibility is $P$=1.02 days and the photometric modulations are caused by some combination of the irradiation effect on the companion and the beaming pattern of the hotspot in the WD's accretion disk.  The second possibility is $P$=2.04 days and the photometric modulations are caused by the ellipsoidal effect with the out-of-round companion alternatively showing its broadside and narrow side as it moves around in the orbit.  The ellipsoidal effects have {\it two} minima and {\it two} maxima during each orbit.  If the ellipsoidal effect is greatly larger than the irradiation and hotspot effects, then the odd-numbered minima will closely match the even-numbered minima, just as observed.

The origin of observed sinusoidal modulations can be tested for similar systems for which radial velocity measures identify the real orbital period.  Here I can quote the cases of four XRBs, with the properties and light curves taken from Schaefer (2025a).  For the famous Cyg X-1 (a BH in a 5.9 day orbit around an O9.7 supergiant), the optical modulations are entirely from the ellipsoidal effect, because the hot and luminous supergiant dominates over the accretion effects.  For the HMXB V884 Sco (a NS in a 3.4 day orbit around an O6 supergiant), the optical modulations are entirely from the ellipsoidal effect, again because the supergiant starlight dominates over the irradiation and the hotspot.  For the LMXB V606 Mon (a BH in a 0.32 day orbit around a K4 main sequence star), the ellipsoidal effect dominates the optical light curve because its normal quiescence has a very low accretion luminosity.  For the prototype-LMXB Sco X-1 (a NS in a 0.787 day orbit around a G-type subgiant star), the irradiation effects are large (due to the extremely high X-ray luminosity), which dominates over the ellipsoidal effects.  These examples show that ellipsoidal modulations are common, while it takes a detailed study to recognize when the ellipsoidal variations dominate or are negligible.  The lesson for FQ Cir is that ellipsoidal modulation is expected, but the ellipsoidal modulation will be apparent only if the irradiation does not dominate.

Both the irradiation and hotspot mechanisms are greatly too weak to be observable in FQ Cir.  For the size of the irradiation effect from a 1.25 $M_{\odot}$ WD with a high accretion of 10$^{-8}$ $M_{\odot}$ yr$^{-1}$, the total accretion energy irradiating the inner edge of the companion for a 1.02 day orbit has a flux that is 2.9\% of the $\sigma T^4$ flux from the B-star alone, and such cannot measurably change the surface brightness.  So the irradiation effect is negligibly small.  For the size of the hotspot effect (and other accretion disk beaming possibilities), for a high accretion of 10$^{-8}$ $M_{\odot}$ yr$^{-1}$ and a hotspot position that is at twice the circularization radius, the total available energy has a luminosity that is 60-millionths of the B-star luminosity.  The flux from the hotspot and other accretion disk structures is greatly too small to measure against the background of the companion star, for any beaming pattern of the light.  So simple energetics demonstrates that the observed photometric modulation cannot be from either the irradiation or hotspot effects.

The ellipsoidal effects must be present in FQ Cir, and their amplitude will only depend on the mass ratio and the orbital inclination, for the case of the B-star anywhere near to filling its Roche lobe.  Leahy \& Leahy (2015) report the calculation of the Roche lobe sizes in the various axes.  For mass ratios anywhere near the known value, the relative radii are constant.  For the FQ Cir case, the distances from the companion star's center (in units of the semi-major axis) are 0.73 to the inner Lagrangian point, 0.62 to the backside near the second Lagrange point, 0.60 to the equator towards and away from the orbital motion, and 0.54 towards the poles above and below the orbital plane.  The full-amplitude of the ellipsoidal modulation is then 0.12 mag when viewed at high inclination.  The observed full-amplitude will be 0.12 mag times the sine of the inclination, being zero for the orbit being seen pole-on.  This amplitude must be present in the FQ Cir binary, with no way around it.

So the ellipsoidal effect must be present and it must dominate greatly over the irradiation and hotspot effects.  With the ellipsoidal effect having two photometric minima each orbit, the orbital period must be exactly twice the photometric period.  So, $P$=2.041738$\pm$0.000036 days.

We are tempted to derive the binary inclination.  For this, we take the 90$\degr$-inclination amplitude to be 0.12 mag and the observed amplitude to be 0.028 mag.  This implies an inclination of 14$\degr$, which is near pole-on.  If the B-star does not fill its Roche lobe, then the companion's shape will be somewhat less out-of-round, and the ellipsoidal amplitude will be smaller.  So we really only have an approximate limit for the inclination of $\gtrsim$14$\degr$.

\subsection{Identity of the Quiescent Counterpart}

The primary mystery of FQ Cir starts with its extremely small nova amplitude of 3.1 mag.  There are only two possible hypotheses for explanation.  The first hypothesis is that the quiescent counterpart is the 14th-mag star, so the nova amplitude is really 3.1 mag.  In this case, the nova light has the usual large amplitude, but such is hidden because the companion star is highly luminous (compared to the usual CVs) so that the combined nova-plus-binary light appears with a small amplitude.  The second hypothesis is that the quiescent nova counterpart is hiding on top of the foreground 14th-mag star.  In this case, the light from the nova binary would have an ordinary amplitude, but the added foreground starlight makes an apparently small amplitude.

We have three high-accuracy measures of the J2000 coordinates of the nova light:  First, Pearce (2022) used the 0.50-m telescope at Siding Spring Observatory to measure 15:24:47.60 $-$60:59:47.3.  Second, Aydi et al. (2022) used the SOAR telescope to measure 15:24:47.63 $-$60:59:47.3 (with respect to UCAC3 stars), with an error bar of $\pm$0.2 arc-seconds.  Third, the {\it Gaia} Transient Name Server\footnote{\url{https://www.wis-tns.org/object/2022nyt}} gives coordinates 15:24:47.610 $-$60:59:47.29, with an uncertainty of 0.055 arc-seconds.  All three positions are closely consistent, and we will adopt the {\it Gaia} position for the nova light.

The 14th-mag star at the nova position has had its position measured by {\it Gaia} for times away from the nova eruption.  This star is Gaia DR3 5875610751015084544, with $G$=13.75, $BP$=13.98, and $RP$=13.25.  This star has J2000 coordinates 15:24:47.611 $-$60:59:47.32.  The 14th-mag star is 0.034 arc-seconds from the nova position.  The positional coincidence is striking, making for a strong case that the real quiescent counterpart is the 14th-mag star.

This positional coincidence can be quantified.  Gaia reports 58 stars within 20'' radius of the nova position.  The probability that a chance coincidence of a field star to within 0.055'' of the nova position is 0.00044 (=1/2280).  This corresponds to a 3.5-sigma confidence level.  So the positional coincidence of the nova and the 14th-mag star is significant.  This astrometric test points strongly to the first hypothesis.

We can now compare the likelihood of the two hypotheses.  Formally, this is by calculating the Bayesian likelihood ratios from hypothesis one to hypothesis two.  We will do this for three evidences; the positional coincidence with the nova light, the presence of a hot/blue/flickering/emission star, and the star being close to filling a Roche lobe.

First, we can ask the likelihood that a star will appear so closely to the nova coordinates?  In the first hypothesis (that the 14th-mag star is the nova counterpart), the positioning of the counterpart will match the {\it Gaia} position.  The counterpart would need to have an eruption amplitude $<$10 mag or so for the counterpart to be detected.  In the grand compilation of measured nova amplitudes (Schaefer 2025b), 33\% (57 out of 172) have amplitudes $<$10 mags.  In the second hypothesis (that the 14th-mag star is a random fore/background star), the likelihood of a random star appearing at the nova position is 0.00044 (see above).  So the ratio of the likelihoods for hypothesis one to hypothesis two is 0.33/0.00044, or 750.  In words, given the evidence that a star appears at the exact location of the nova eruption, it is likely for the nova counterpart to appear at the nova position, while it is unlikely for a random star to appear at the nova position, with the likelihood ratio of 750 providing strong evidence for the first hypothesis.

Second, we can ask for the likelihood that the 14th-mag star observed at the nova position has the observed properties of being hot, blue, flickering, and with emission lines?  In the first hypothesis where the 14th-mag star is the nova quiescent counterpart, the likelihood of the counterpart being a hot/blue/flickering/emission star is near to unity.  In the second hypothesis where the 14th-mag star is a random fore/background source, the likelihood of a random Galactic star being hot, blue, flickering, and emission-lined is extremely small.  In the exhaustive and magnitude-limited {\it Yale Bright Star Catalog} (Hoffleit 1982), the fraction of hot blue stars (i.e., the O, B, and Wolf-Rayet stars) is 20\%.  As quoted in Section 2.2, a strong limit can be placed on the frequency of hot blue stars having substantial flares or trends, and that limit is $<$0.4\%, which is certainly an high over-estimate.  The hot blue stars with emission lines are a similar population.  So the overall probability of a random galactic star in a magnitude-limited sample being a hot/blue/flickering/emission star is roughly $<$$\frac{1}{1250}$.  Given the evidence that the 14th-mag star is hot, blue, flickering, with emission lines, the likelihood ratio between the two hypotheses is $>$1250 in favor of the 14th-mag star being the nova system.

Third, we can ask for the likelihood that the 14th-mag star nearly fills its Roche lobe in a binary orbit?  That is, the 2.04 day orbital period and the measured radius of the 14th-mag star make for it to nearly fill its Roche lobe, which would make it an interacting binary star with accretion.  In the first hypothesis, the nova counterpart must be an interacting binary so as to feed the accretion onto the WD.  So the likelihood is near unity for seeing a near interacting binary in the case that the 14th-mag star is the nova.  In the second hypothesis, a random star has only near-zero chance of being an interacting binary.  This can be quantified from the {\it Yale Bright Star Catalog} (Hoffleit 1982), where zero out of 9096 stars are interacting binaries.  That is, the likelihood is $<$$\frac{1}{9096}$ that a random star in a magnitude limited sample will have the star nearly filling its Roche lobe.  Given the evidence that the 14th-mag star is nearly filling it Roche lobe, the likelihood ratio between the two hypotheses is $>$9096.

The three evidences are independent, so the three likelihood ratios should be multiplied together so as to get a final likelihood ratio between the hypotheses for explaining the evidences.  We have ratios of 750 for the star being at the nova coordinates, $>$1250 for the evidence that the star is hot/blue/flickering/emission, and $>$9096 that the star is nearly filling it Roche lobe.  Multiplied together, the combined likelihood ratio is $>$8$\times$10$^9$.  This is large enough so that we take the 14th-mag star to be the true counterpart of the nova eruption.  From now on, we take the 14th-mag star as the companion star in the nova binary.

\subsection{Extinction}

First, for the quiescent counterpart, the {\it Gaia} satellite reports that $E(BP-RP)$ is 1.063$\pm$0.002 mag.  With the color transformations from Table 3 of Jordi et al. (2010), this gives $E(B-V)$=0.69, with small error bars.

Second, Schlafly \& Finkbeiner (2011) give the $E(B-V)$ for the entire column through our galaxy to be 0.629$\pm$0.028 mag.  For any reasonable distance, FQ Cir is far above nearly all of the extinction in the plane of our Milky Way.  So the extinction to FQ is close to 0.629$\pm$0.028 mag.

Third, the peak had $B-V$ color of 0.68$\pm$0.03 mag.  The intrinsic color of novae at peak is $+$0.11$\pm$0.04 mag (Schaefer 2022b).  So FQ Cir has $E(B-V)$=0.57$\pm$0.05.

From these three measure, $E(B-V)$=0.64$\pm$0.05.  With the usual ratio of 3.1 appropriate for the nearby regions of our galaxy, the total visual extinction is 2.0 mag.

\subsection{Distance}

The {\it Gaia} parallax to the nova counterpart is 0.113$\pm$0.019 milli-arc-seconds.  The distance calculated by the {\it Gaia} Team should not be used, because they adopted an inappropriate prior for the distance distribution of the parent population, and because they have not included additional independent measures of the nova distance.  In particular, FQ Cir is in the nova disk population, so the appropriate prior is for an exponential distribution with a scale height of 150 pc (Schaefer 2022b).  Further, additional information on the FQ Cir peak can be used as a prior for the Bayesian parallax calculation.  That is, the observed peak magnitude is $V_{\rm peak}$=10.9, while novae have the overall average peak absolute magnitude of $-$7.45 with an RMS scatter of $\pm$1.4 mag.  With the correct Bayesian calculation as in Schaefer (2022b), the distance to FQ Cir is 8800$\pm$1500 pc.

\subsection{Absolute Magnitudes}

The nova peaked at $V_{\rm peak}$=10.9$\pm$0.1, so with $D$=8800$\pm$1500 pc and $E(B-V)$=0.64$\pm$0.05, the $V$-band absolute magnitude is $-$5.8$\pm$0.4.  This qualifies FQ Cir as a  fast-faint nova (Kasliwal et al. 2011, see Figure 13), but this is just near the edge of of the distribution for ordinary novae (Schaefer 2022b, see Figure 4).

The quiescent counterpart has $V$=14.0$\pm$0.1.  This translates to an absolute magnitude of $-$2.7$\pm$0.4 mag.  This makes FQ Cir as the most luminous CV in quiescence (Dubus et al. 2018, Patterson et al. 2022, Schaefer 2025b).  This measure will be critical for calculating the blackbody radius of the companion (see Section 2.14 and Figure 8).

\subsection{Colors and Spectral Energy Distribution}

\begin{figure}
	\includegraphics[width=1.01\columnwidth]{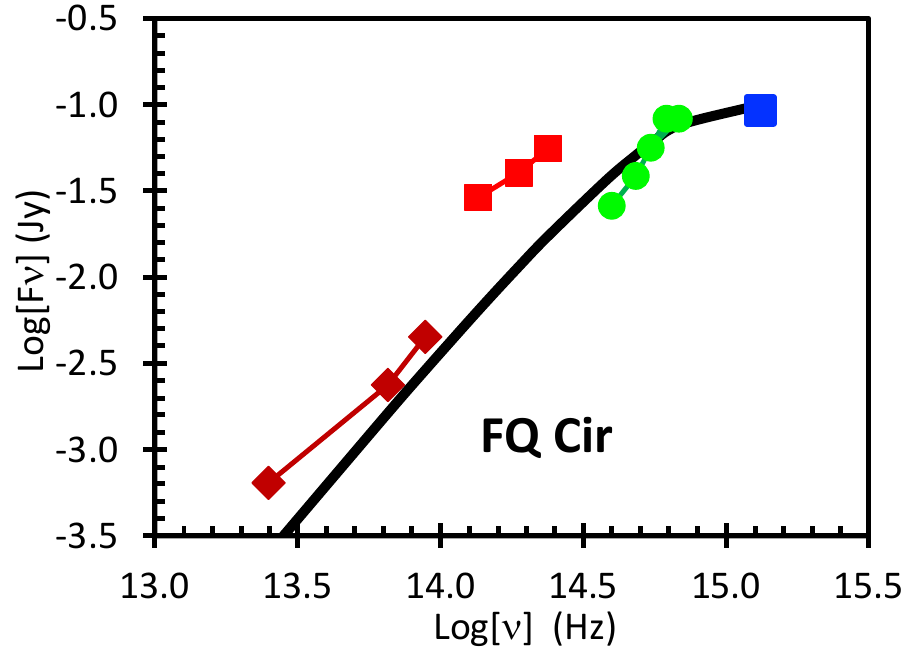}
    \caption{Spectral energy distribution for the quiescent counterpart of FQ Cir.  The flux (in units of Jansky) is plotted versus the central photon frequency, $\nu$.  These data are from the infrared with {\it WISE} (burnt-red diamonds), the near-infrared with 2MASS (red squares), the optical with APASS (green circles), and the ultraviolet with {\it Galex} (the blue square).  The data sets do not follow any one smooth curve due to the known variability and the data sets being taken in different years.  Nevertheless, there is a clear Rayleigh-Jeans slope throughout the infrared, and there is a turnover towards the ultraviolet.  The Rayleigh-Jeans slope shows that the SED is dominated by a blackbody, and the blue-turnover shows that the surface temperature is 20,000$\pm$3000 K.   Further, we see no evidence of any accretion disk or dust shell or decretion disk, and we do not expect any because the Be-star dominates over any plausible emission.}
\end{figure}

The observed magnitudes at peak are best taken from the AAVSO light curve\footnote{see \url{https://www.aavso.org/LCGv2/}}.  Within hours of peak, the first fully calibrated CCD measures are $B$=12.08 and $V$=11.40 from A. Pearce, plus $R$=10.42 and $I$=9.69 from F. Romanov.  The extinction corrected colors are $B-V$=$+$0.04, $V-R$=$+$0.47, $R-I$=$+$0.17.

For the best colors in quiescence, we can do no better than the average magnitudes from the AAVSO Photometric All-Sky Survey\footnote{\url{https://www.aavso.org/apass}} (APASS).  These magnitudes are on eight nights from 2011--2014, with definitive calibrations against the Landolt standard stars.  We have $B$=14.42, $V$=13.96, $g$=14.10, $r$=13.73, and $i$=13.62.  The $B$ and $V$ magnitudes show RMS variability of 0.11 mag over 6 nights.  The extinction corrected $B-V$ is $-$0.18, with the variability and extinction correction making for an uncertainty of $\pm$0.08 mag.  This is a very blue color.  

The spectral energy distribution (SED) for the counterpart can be constructed from the {\it GALEX} near-ultraviolet measure, the APASS visible magnitudes, the 2MASS near-infrared bands, and from the {\it WISE} infrared measures.  The {\it GALEX} $NUV$ magnitude is 16.57, centered on 2271~\AA.  The 2MASS magnitudes are $J$=11.71, $H$=11.42, and $K$=11.12.  The {\it WISE} magnitudes are $Wise1$=12.25, $Wise2$=12.27, $Wise3$=11.81, and $Wise4$=8.84.  These magnitudes are extinction corrected and converted to flux in the units of Jansky.  Figure 5 shows a plot of the flux versus the central wavelength.

The FQ Cir SED shows a steep rise from the infrared points from {\it WISE} (in the lower left in Figure 5), to the {\it GALEX} ultraviolet (in the upper right).  For comparison, a 20,000 K blackbody is plotted.  The 2MASS near-infrared magnitudes (the red squares) are well above the blackbody, while the APASS optical magnitudes (the green circles) are on-or-below the blackbody.  These differences are a result of the known variability, for which each of the data sets were taken in different years.  With this variability, no blackbody fit can have any high accuracy.  Nevertheless, the SED does show a Rayleigh-Jeans slope in the infrared and a turnover towards the ultraviolet.  With the Rayleigh-Jeans slope, we know that we are dealing with a blackbody.  With the turnover towards the blue, we know that the counterpart has a temperature something like 20,000 K. Given the variability, the temperature might be anywhere from 17,000--23,000 K.

\subsection{Optical Spectrum of the Counterpart}

\begin{table}
	\centering
	\caption{Journal of FQ Cir spectra}
	\begin{tabular}{llrll}
		\hline
		Spectrum   &  Aperture  &  R  &   Date (JD)   &  Range ($\AA$) \\
		\hline
MB1	&	0.3-m	&	614	&	2459757.446	&	3800--7500	\\
MB2	&	0.3-m	&	608	&	2459758.313	&	3800--7500	\\
MB3	&	0.3-m	&	669	&	2459762.490	&	3800--7500	\\
MB4	&	0.3-m	&	429	&	2459764.470	&	3800--7500	\\
MB5	&	0.5-m	&	500	&	2460874.996	&	3800--7500	\\
MB6	&	0.5-m	&	500	&	2460901.919	&	3800--7500	\\
MB7	&	0.5-m	&	500	&	2460910.876	&	3800--7500	\\
MB8	&	0.5-m	&	500	&	2460912.886	&	3800--7500	\\
SALT1	&	9.2-m	&	14,000	&	2460921.849	&	3763--8790	\\
SALT2	&	9.2-m	&	14,000	&	2460923.838	&	3763--8790	\\
SS1	&	2.3-m	&	7000	&	2460948.400	&	4184--7024	\\
SS2	&	2.3-m	&	7000	&	2460950.402	&	4184--7024	\\
SS3	&	2.3-m	&	7000	&	2460952.402	&	4184--7024	\\
SS4	&	2.3-m	&	7000	&	2460961.409	&	4184--7024	\\
		\hline
	\end{tabular}		
\end{table}

\begin{figure*}
	\includegraphics[width=2.1\columnwidth]{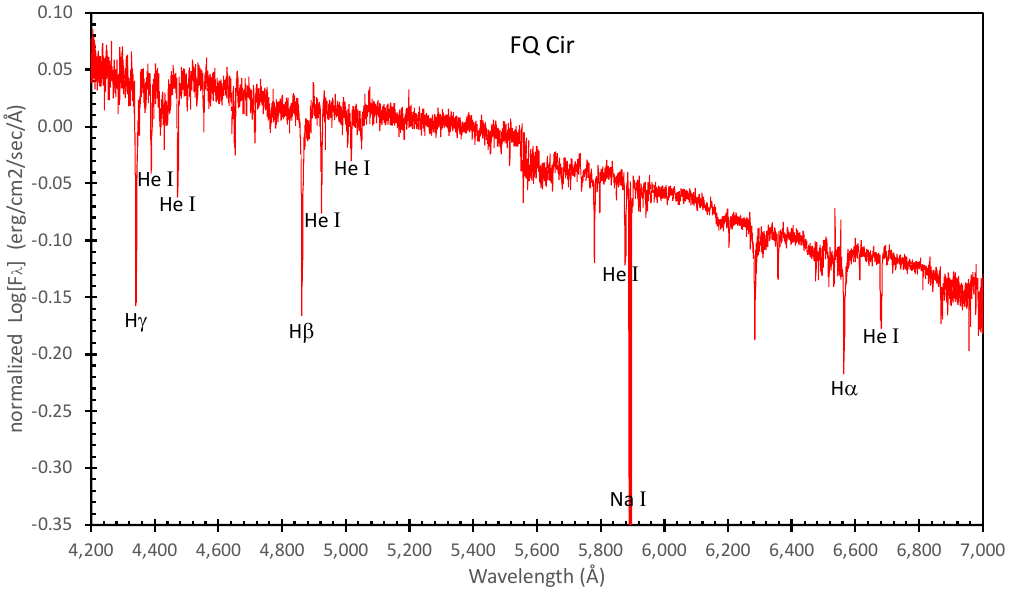}
    \caption{Spectra for FQ Cir in quiescence.  The Siding Spring spectra (SS1--SS4) have been calibrated into physical units, normalized with the logarithms, then averaged.  The small continuum step at 5548~\AA~is instrumental.  The interstellar sodium doublet (centered at 5890.3 and 5896.3~\AA) extends below the plot down to -0.8 and -0.6 in the logarithmic units.  All of the prominent lines are hydrogen Balmer lines or He I lines.  Unlabeled lines include the interstellar PAH band around 4430~\AA, and the diffuse interstellar bands at 5780 and 6283~\AA.  This spectrum demonstrates that the companion is a normal B1 V star.  There is no emission above the continuum.  But the H$\alpha$ line is not as deep as the H$\beta$ line, and this can only be because there is substantial emission partially filling in the H$\alpha$ profile.   }
\end{figure*}

\begin{figure*}
	\includegraphics[width=2.1\columnwidth]{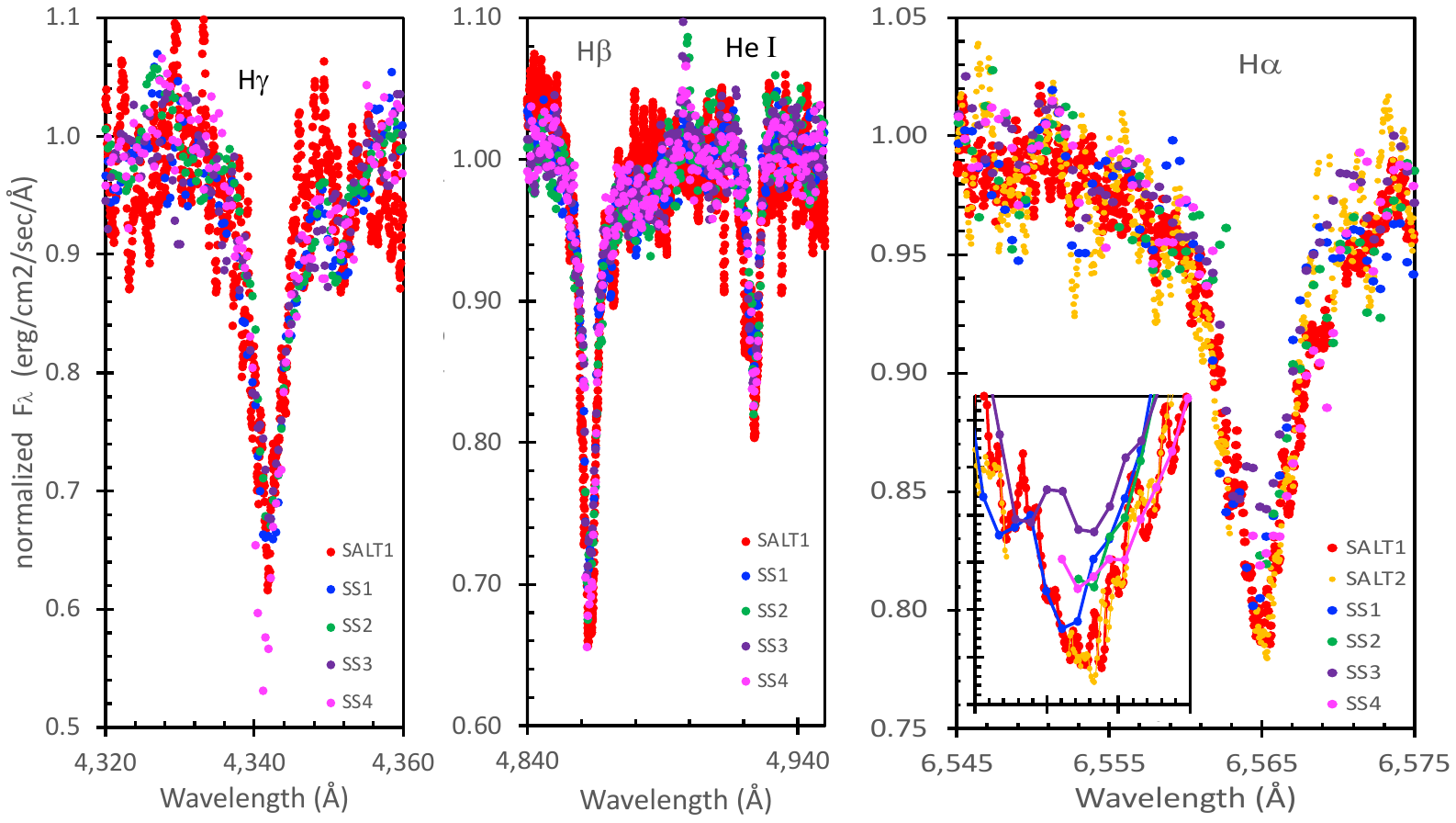}
    \caption{Balmer line profiles.  These line profiles for H$\gamma$, H$\beta$, He I 4921~\AA, and H$\alpha$ have smooth and deep absorption, display no obvious emission lines, and are largely constant from night-to-night.  On the face of it, this means that the B-star has no more than small emissions.  Nevertheless, three confident observations indicate that the B-star does indeed have substantial emission, and merits the ``(e)'' in the spectral classification, denoting a weak Be star.  The first evidence is that the depth of the H$\alpha$ absorption (with a typical minimum at 0.8 times the continuum level) is shallower than H$\beta$ (typically falling to 0.7 time the continuum), which in turn is shallower than H$\gamma$ (typically falling to 0.65 times the continuum), with this being the opposite of the case for normal B-stars, with the only explanation being the hydrogen emission is filling the H$\alpha$ profile and to a lesser extent lightly filling the H$\beta$ profile.  The second evidence is that the H$\alpha$ profiles are variable from night-to-night (see the inset), and this can only be due to variable emission fill-in.  The third evidence is that the SS3 and SS4 spectra have a flat bottomed H$\alpha$ profile well above the usual sharp cores (see the inset on the rightmost panel), and this can only be due to the partial filling-in of the line by emission.  While there must be emission like in a Be star, the relatively low strength of the emission is caused by the decretion disk being truncated by the Roche lobe.   }
\end{figure*}

Love (2022) has the only publicly available spectra of FQ Cir, with these appearing in the Astronomical Ring for Amateur Spectroscopy (ARAS) database\footnote{\url{https://aras-database.github.io/database/novacir2022.html}}.  There are four eruption spectra, from 2022 June 26 to July 3, on days +2 to +9 after the start of the eruption.  These spectra were made with a 30 cm telescope in Martinborough, New Zealand, each from 3800--7500~\AA, with resolving power $R$ around 600.  The four eruption spectra are labeled MB1 through to MB4 (see the journal in Table 1).  These show the usual broad Balmer lines (with H$\alpha$ FWHM 2600 km s$^{-1}$) and He I lines.  These provide the proof that FQ Cir is just a normal CN, of the He/N classification.

We have taken 10 spectra in post-eruption quiescence, as summarized in Table 1.  The four spectra from Martinborough are labeled MB5 to MB8, with dates 2025 July 18 to August 25.  These four spectra were made with a 0.5-m telescope, also in Martinborough.  The two spectra taken with the Southern African Large Telescope (SALT) 9.2-m telescope on 2025 September 3 and 5 are labeled as SALT1 and SALT2.  These SALT spectra were obtained with the HRS dual-beam echelle spectrograph, as processed with the SALT primary data pipeline (Kotze et al. 2025) and the HRS data reduction pipeline (Kniazev, Gvaramadze, \& Berdnikov 2016).  The four spectra taken with the WiFeS spectrograph (Dopita et al. 2010) on the ANU 2.3-m at Siding Spring Observatory from 2025 September 9 to October 12 are labeled SS1 to SS4.  Figure 6 plots the spectra from the average for SS1--SS4.  Figure 7 overplots the individual spectra in the region of the first three Balmer lines.

The MB5--MB8 spectra were only expected to be a reconnaissance of the optical spectrum, so the signal-to-noise ratio is substantially poorer than possible with a big telescope.  This circumstance means that the weak lines used to formally classify the spectral class and luminosity class are not detected.  Nevertheless, the spectra are adequate for seeing the nature of the quiescent nova.  In particular, we saw that the companion was an early B-class star, likely on the main sequence.  Importantly, we saw no emission above the continuum.  Further, these four spectra are useful for demonstrating that the base spectra of the B-star do not have any large variability, for example, the appearance and disappearance of emission lines.  These first-look spectra guided our early analysis, as well as our planning of later spectral observations.

The Siding Spring spectra are adequate to define the spectral class, making allowance for their quite low signal to noise.  For comparisons versus the standard stars, the closest match is to $\omega^1$ Sco, a normal B1 V star (see figures 3.5 and 4.1 of Gray \& Corbally 2009).  With the comparison to $\omega^1$ Sco, the FQ Cir Balmer line widths are slightly larger, indicating a larger $v\sin i$, with this denoted by the flag ``(n)''.  For the temperature classification, the primary indicator is the ratio of the hydrogen to helium lines.  To avoid the effects of the hydrogen cores being filled in with emission to some extent, it is best to use the absolute strengths of the He I lines in relation to overall features.  Since He I reaches a maximum strength at B2, then B1 and B3 are going to have about the same ratio of He I. Other features, specifically the ionized lines, put a star hotter or cooler than B2 (see fig 4.1 in Gray \& Corbally 2009), showing FQ Cir is on the hotter side. These considerations give B1 type.  From the overall features, the spectral class is probably not earlier than B0.5 or later than B2.  In all, the best measure of the spectral class is B1, with a possible range from B0.5 to B2.

For the luminosity class, the line ratios give a definite answer.  The ratio of the O II and C III lines near 4650~\AA~to the helium lines at 4713~\AA~and 4388~\AA~indicate that the luminosity class is definitely V.  The ratio of the Si III line at 4552~\AA~to the helium lines at 4713~\AA~and 4388~\AA~corroborates this, as does the absence of any absorption for the Mg II 4481~\AA~ line.  So the companion is on the main sequence, denoted with ``V''.

The line profiles for the three Balmer lines (H$\gamma$, H$\beta$, and H$\alpha$) are displayed in Figure 7.  The Balmer lines have roughly the expected shape for a B-star.  These profiles leave little room for any large emission inside the absorption lines.  Nevertheless, there must be some filling in of the Balmer absorption lines.  The first evidence is that the absorption depth of H$\alpha$ (running to 0.80 times the continuum) is shallower than the depth of H$\beta$ (running to 0.70 times the continuum) and is shallower than the depth of H$\gamma$ (around 0.65 times the continuum).  In early B-type stars with no emission, H$\alpha$ should be deeper than H$\beta$, which should be deeper than H$\gamma$.  The only way to reverse this usual `Balmer decrement' is to have emission like in a Be star, because the emission is always by far the brightest in H$\alpha$ and relatively faint in H$\gamma$ (Catanzaro 2013).  The second evidence is that the detailed profile in H$\alpha$ is variable, see the inset in Figure 7.  The photosphere cannot change night-to-night or in the manner observed, so the changing profile can only be caused by variable emission lines.  Variable emission lines are one of the hallmarks of Be stars (Rivinius, Carciofi, \& Martayan 2013).  The third evidence is that some of the H$\alpha$ profiles (especially for SS3 and SS4) are flat bottomed.  Without emission inside the H$\alpha$ line, the profile would have a sharp minimum.  The flat profile demonstrates that some substantial emission is present inside the H$\alpha$ line.  With these evidences, we see that FQ Cir does have Balmer emission, and this requires the classification to have an ``(e)'' added to the end.  That is, FQ Cir is a Be star, albeit a weak Be star.

The Balmer line widths show slight rotational broadening, in comparison with slow rotating B1 V stars, and this is indicated in the classification with "(n)".  (This broadening is not as much as ``n", while ``nn" means the largest rotational broadening.)  Given that we already think that the inclination is small (due to the small amplitude of the ellipsoidal modulation), this would require a fast rotating companion to produce the observed slight rotational broadening.

The full spectral classification of the companion is B1 V(n)(e).  That is to say, the companion star is a B1 star on the main sequence, showing slight rotational broadening, plus is a weak and variable Be emission line star.

\subsection{White Dwarf Mass}

The nova counterpart must have a WD so as to produce its normal nova eruption.  We can get a reasonably-good estimate of the WD mass ($M_{\rm WD}$) from four measures of the nova light:

First, the He/N classification requires that the WD be relatively high in mass.  Schaefer (2025b) uses the measures of $M_{\rm WD}$ for 152 novae with spectral classification, with the result consistent that all the He/N novae have WD masses $>$1.15 $M_{\odot}$.

Second and third, the decline rate of nova is closely correlated with the WD mass.  The method of Shara et al. (2018) can work on the $t_2$=2 days measure, for $M_{\rm WD}$=1.30 $M_{\odot}$.  (For this method, the appropriate amplitude input is unclear, but this makes for little uncertainty in the mass estimate.)  Schaefer (2025b) uses the $t_3$$\le$11 days to estimate $>$1.31 $M_{\odot}$.  

Fourth, the FWHM of 2600 km s$^{-1}$ is well above average for novae, pointing to a relatively high mass.  Schaefer (2025b) has quantified the FWHM versus $M_{\rm WD}$ relation with 127 novae, where the observed H$\alpha$ FWHM corresponds to 1.20$\pm$0.10 $M_{\odot}$.

From these four measures, we take the FQ Cir WD mass to be 1.25$\pm$0.10 $M_{\odot}$.

\subsection{Companion Surface Temperature}

We have five ways of getting the surface temperature of the optical star in the counterpart:

First, the optical spectra return a spectral class of B1, with a range of B0.5 to B2.  This corresponds (see Table B.3 in Gray \& Corbally 2009) to an effective temperature of 24,500 K, with a range of 19500 to 26750 K.

Second, the SED in Figure 5 has a turnover into the ultraviolet.  This corresponds to a temperature of 20,000$\pm$3000 K.

Third, {\it Gaia} directly reports an effective temperature of 22483$\pm$120 K.  We are concerned that the quoted error bar is unreasonably small, likely only representing the propagation of measurement errors without including any systematic errors.  Specifically, the variability of the continuum (from flickering) and the comings and goings of the emission in the lines might make the real error bars be substantially larger than the quoted $\pm$120 K.

Fourth, the extinction-corrected color in quiescence is $B-V$ of $-$0.18$\pm$0.08.  For getting the surface temperature, we use the calibrations in {\it Allen's Astrophysical Quantities} (Cox 2000, 4th edition, Table 15.7).  The observed color corresponds to a B4 star, with a range of B1--B8.  This also corresponds to a surface temperature of 17100 K, with a range of  11,400--25,000 K.

Fifth, we measured the absolute magnitude in the $V$-band to be $-$2.7$\pm$0.4.  With interpolation between Tables B.1 and B.3 of Gray \& Corbally, this translates to a temperature of 20500 with a range of 18500 to 22500.  So our measure of the absolute magnitude in quiescence gives a surface temperature of 20500$\pm$2000 K.

Given these five measures, we judge that the surface temperature is 21,500$\pm$1500 K.  Or, if we accept the small error bars on the {\it Gaia} temperature measure, the surface temperature is near 22,483 K.

With the analysis as in Sections 2.13 and 2.14, we see inconsistencies between the measured blackbody radius and the main sequence radius for temperatures below roughly 21,000 K.  With this, the possible temperature range would only be 21,000 to 23,000 K.  This can be represented as 22,000$\pm$1000 K.  Or, if we accept the small error bars on the {\it Gaia} temperature measure, the surface temperature is near 22,483 K.

\subsection{Companion Mass}

\begin{figure}
	\includegraphics[width=1.01\columnwidth]{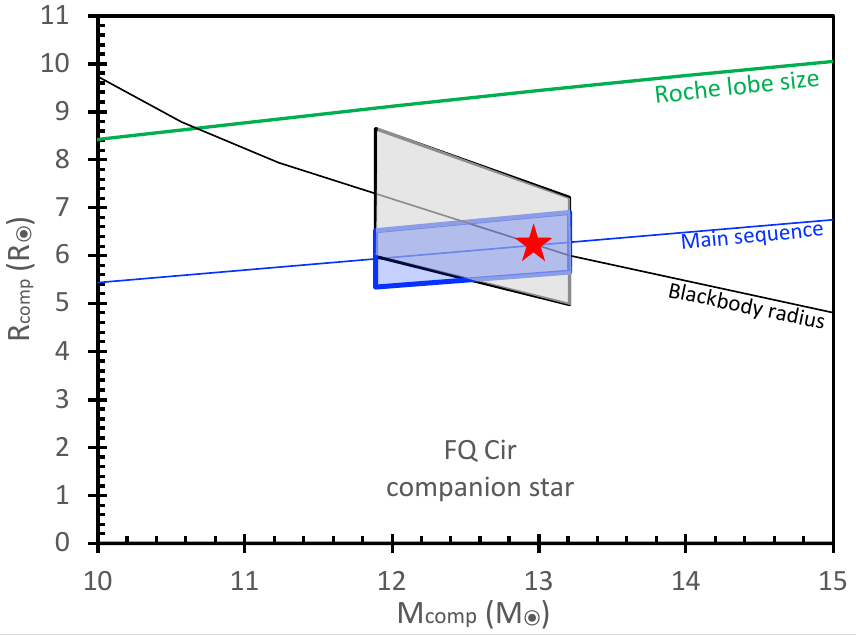}
    \caption{Radius/Mass constraints for the companion star.  The main sequence for B stars is represented by the blue line, with the blue box representing the allowed range for the companion star.  The measurement of the blackbody radius is represented by the thin black line, with the gray box showing the allowed range for the companion star.  The most likely position (where the two centerlines cross) is shown with a red star.  This gives $M_{\rm comp}$=13.0$^{+0.2}_{-0.5}$ $M_{\odot}$ and $R_{\rm comp}$=6.2$\pm$0.2 $R_{\odot}$.  The green curve across the top shows the Roche lobe radius.  The size of the companion star is far inside the Roche lobe.  The decretion disk covers radii 6.2 to 9.5 $R_{\odot}$, where it is truncated by the Roche lobe.  }
\end{figure}

We do not have any radial velocity curve to give us the stellar masses.  Instead, the best information comes from the knowledge that the companion is on the main sequence and with a temperature of 22,000$\pm$1000 K.  For the main sequence, we use the mass-temperature relation is given in Tables 15.7 and 15.8 of Cox (2000).  For our best-estimate temperature of 22,000 K, the companion mass is 12.55 $M_{\odot}$.  The acceptable range is for temperatures of 21,000--23,000, which corresponds to main sequence masses of 11.89--13.21 $M_{\odot}$.  Alternatively, if we accept the small error bars on the {\it Gaia} temperature of 22483$\pm$120 K, then we get a mass close to 12.87 $M_{\odot}$

With the analysis in the next section, we combine the two independent constraints of our measured blackbody radius and the main sequence.  As shown in Figure 8, the two constraints formally cross at 13.0 $M_{\odot}$, corresponding to a main sequence temperature of 22,700 K.  The allowed region for the main sequence is the blue shaded are, with the highest probability close to the blue main sequence line.  The vertical height of this box is $\pm$10\%, to represent the real scatter of stars in the main sequence.  The allowed region from the blackbody radius measure is the gray shaded region, with the upper edge for absolute magnitude of -3.1 mag, and the lower edge for absolute magnitude of -2.3 mag.  For the blackbody radius measure, the highest probability is along the black curve (with absolute magnitude -2.7 mag).  Both error regions are bounded on the left at mass 11.89 (corresponding to temperature 21,000 K) and on the right at mass 13.21 (corresponding to temperature 23,000 K.  To get the joint probability distribution from the two measures (that the companion is on the main sequence and has absolute magnitude of -2.7$\pm$0.4), we have to multiply together the individual independent probability distributions.  On the left side with temperatures below 21,000 K, the two curve have diverged sufficiently that the joint probability is low.  This is the justification in the previous section for changing the lower boundary from a temperature of 20,000 K to a temperature of 21,000 K.  Further, this means that left side of the joint region in Figure 8 will have lower probability than the right side, thus shifting our best estimate mass somewhat to the right of the middle of the mass range.  The most probable companion mass is for the position in Figure 8 where the two centerlines cross.  This is at 13.0 $M_{\odot}$.  The error bars are somewhat asymmetric, so we have 13.0$^{+0.2}_{-0.5}$ $M_{\odot}$.

Alternatively, if we accept the small error bars on the {\it Gaia} temperature, then the companion has a radius close to 12.9 $M_{\odot}$.  What we are seeing is a close agreement between the $M_{\rm comp}$ derived from temperatures quoted by {\it Gaia}, the mass derived from our independent temperatures not using {\it Gaia}, and the mass derived from the formal crossing of the two constraints.  And all of these solutions have effectively the same stellar radius for the companion star.

\subsection{Companion Radius}

The radius of the companion star can be measured with four methods:

First, we can turn the observed spectral class into a radius, as based on the reasonable assumption that the companion is a normal B type main sequence star.  With the canonical calibration from Cox (2000, Table 15.8), the B1 V classification corresponds to 6.5 $R_{\odot}$.  The full range of B0.5--B2 V corresponds to 5.7--6.9 $R_{\odot}$.  This does not include the scatter of stars around the main sequence.

Second, a better method is to use all of our constraints on the surface temperature, yielding 22,000$\pm$1000 K, to place the companion onto the main sequence.  This is done in Figure 8, with the blue box.  The best value for the companion radius is 6.1 $R_{\odot}$.  The range is from 5.3 to 6.9 $R_{\odot}$.

Third, we now have enough information to calculate the blackbody radius of the companion star.  The application to FQ Cir is justified by the blackbody SED, with a long Rayleigh-Jeans section and a turnover to the blue (Fig. 5).  This calculation is simple, just equating the luminosity ($L$) from the Stefan-Boltzmann Law and from the bolometric absolute magnitude, all scaled to our Sun.  This gives the equation
\begin{equation}
\frac{4 \pi R_{\rm comp}^2 \sigma T_{\rm comp}^4}{4 \pi R_{\odot}^2 \sigma T_{\odot}^4} = \frac{L_{\rm comp}}{L_{\odot}} = \frac{10^{-0.4 M_{\rm bol,comp}}}{10^{-0.4 M_{\rm bol,\odot}}}.
\end{equation}
We have already found $T_{\rm comp}$=22,000$\pm$1000 K (see Section 2.12), while $T_{\odot}$ is 5777 K, and $M_{\rm bol,\odot}$=4.74.  The bolometric correction for the temperature of the companion is $-$2.45 (see Table 15.7 of Cox 2000), the absolute $V$ magnitude is $-$2.7$\pm$0.4 mag, so $M_{\rm bol,comp}$ is $-$5.15$\pm$0.40.  Equation 1 can then be solved for $R_{\rm comp}$ in units of $R_{\odot}$.  This gives the best estimate for the companion radius of 6.55 $R_{\odot}$.  For every $M_{\rm comp}$ and its corresponding temperature, the basic solution for the blackbody radius is displayed as the thin black curve in Figure 8.  The region that fits the blackbody radius with uncertainties is shown as a gray box.  The gray box is bounded on the left by $T$=21,000 K (for $M_{\rm comp}$=11.9 $M_{\odot}$), on the right by $T$=23,000 K (for $M_{\rm comp}$=13.2 $M_{\odot}$), on the top by $M_V$ of $-$3.1, and on the bottom by $M_V$ of $-$2.3.
 
Fourth, the orbital period can be used to calculate the companion star's Roche lobe radius, as a function of the companion star's mass.  With this, Kepler's Law gives the semimajor axis, and then the radius of the Roche lobe for the companion star.  The companion star radius must equal to the Roche lobe size (if RLOF is in operation), or somewhat smaller radius (if some other accretion mode is in operation).  If $M_{\rm comp}$ is 13 $M_{\odot}$ then the Roche lobe is 9.5 $R_{\odot}$.  The function of the Roche lobe radius versus the companion mass is plotted in Figure 8.  The companion need not fill the Roche lobe, so the curve in Figure 8 is an upper limit.

Figure 8 displays our two essential constraints on the mass and radius, with the gray box for the measure that $M_V$ is -2.7$\pm$0.4, and with the blue box for the measure that the companion star is on the main sequence.  The center lines for these two relations cross at 13.0 $M_{\odot}$ and 6.2 $R_{\odot}$.  Simplistically, the two joint constraints are given by the intersection of the gray and blue boxes.  A better representation of the joint constraints is probability from the normalized product of our two independent constraints, where each constraint has the probability maximal along the centerline and falling off to low values near the box edges.  The maximum probability is where the two centerlines cross.  For temperatures under 21,000 K (to the left of the box edges), the joint probability is unacceptably low (because the main sequence centerline and the blackbody radius centerline have diverged substantially), justifying the raising of the minimum temperature to 21,000 K, as done in Section 2.12.  With the divergence of the centerlines for masses under 12.5 $M_{\odot}$ (i.e., near the middle between the right and left sides), the joint probability is unlikely.  The most likely probability is for masses 12.5--13.2 $M_{\odot}$ and radii 6.0--6.4 $R_{\odot}$.  We represent this result as $M_{\rm comp}$=13.0$^{+0.2}_{-0.5}$ $M_{\odot}$ and $R_{\rm comp}$=6.2$\pm$0.2 $R_{\odot}$.  This is our final best-estimate measures of the mass and radius of the companion star.

If we take the {\it Gaia} error bar on the temperature at face value, then we would have said that the companion mass is near 12.9 $M_{\odot}$ and the companion radius is 6.2$\pm$0.2 $R_{\odot}$.  This is close to our final best-estimate.

The stellar radius of the companion has importance because it determines whether FQ Cir is in RLOF, and if not, then it determines the available size for the decretion disk.  For the question of RLOF, we have a clear answer, with the companion far inside the Roche lobe.  This can be seen in Figure 8, with the acceptable zone always lying far below the Roche lobe size.  Numerically, the star size (6.2$\pm$0.2 $R_{\odot}$) is greatly smaller than the Roche lobe (9.5 $R_{\odot}$ for a 13 solar-mass companion).  So the large required accretion (needed to power the nova eruption) cannot come from RLOF.  (This is as expected, because RLOF in a binary with a high mass ratio must have a well-known dynamical instability that produces exponentially-rising runaway accretion with a timescale or order one century, and such is not seen in the long-term light curve of Figure 1.)  The accretion comes from the decretion disk.  This disk can only cover the radius range 6.2 to 9.5 $R_{\odot}$, for a radius range of a mere 3.3 $R_{\odot}$.  This is substantially smaller than the decretion disks around most Be stars, which extend out to 20--40 $R_{\odot}$ (Rivinius, Carciofi, \& Martayan 2013).  The small size is forced due to the disk being truncated by the Roche lobe.  This small size is a partial explanation for why FQ Cir is a {\it weak} Be star.

The situation of FQ Cir, a rapidly rotating early-B star inside its Roche lobe, means that the companion is a Be star with a decretion disk.  We further know this because the companion has large-amplitude flares/flickering/trends that are unique to Be stars.  And we see the decretion disk by its Balmer emission.  Decretion disks consist of out-spiraling gas, which must ultimately pass through the Roche lobe, and then it can only fall onto the WD through an accretion disk.  Thus, we see FQ Cir as accreting in a new mode, with gas moving from the decretion disk around the companion star to the accretion disk around the WD.

\subsection{Companion Rotation Velocity}

What is the equatorial rotation velocity ($v_{\rm rot}$) of the companion star?  We have five ways get some measure of $v_{\rm rot}$:

First, in comparison with the B1 V standard star $\omega^1$ Sco  (Gray \& Corbally 2009, figure 4.1), the Balmer lines are somewhat broader, due to the FQ Cir companion star showing rotational broadening.  This is what leads to the  ``(n)'' in the classification.  Given that the binary inclination is low ($i$$\gtrsim$14$\degr$), the underlying equatorial rotation velocity ($v_{\rm rot}$) must be large.

Second, the nova in quiescence displays flickering, flares, and trends on all timescale, and such is universal for Be stars, while none of the non-Be stars show such variability.  So the companion is a Be star.  And Be stars all are fast rotators, with the variability caused by effects of the fast rotation.  So the FQ Cir companion has a fast $v_{\rm rot}$.

Third, the companion is observed to have the emission lines of a Be star, and this can only come from the decretion disk thrown off by the fast rotation.  That is, all Be stars show emission only because the fast equatorial rotation has helped to throw off a decretion disk, so the observed existence of the decretion disk requires fast $v_{\rm rot}$.

Fourth, in principle, the observed widths of the Balmer lines should provide a measure of the rotation.  The observed width of the Balmer absorption lines is a convolution of the pressure broadening and the rotational broadening as $v_{\rm rot}$$\sin i$.  Further, emission within the Balmer line profiles will distort the measured widths.  In practice, with variable emission added to the line profiles, with noisy spectra, and with the edge of the wings being ill-defined, we have no good measure of line widths that can be accurately related to $v_{\rm rot}$$\sin i$.  Further, small changes in the spectral class will change the intrinsic width of the spectrum, and this can mimic rotational broadening.  Here, we can report on the half-width zero-intensity (HWZI) and the full-width half-maximum (FWHM) of the Balmer profiles, where in a schematic way, these should equal $v_{\rm rot}$$\sin i$ (after the intrinsic line width is somehow accounted for).  For the profiles in Figure 7, the FWHM velocities for the H$\alpha$, H$\beta$, H$\gamma$ and He I 4921~\AA~lines are 250, 320, 340, and 270 km/s respectively.  For the same lines, the HWZI velocities are 280, 390, 350, and 320 km/s respectively.  These are imperfect measures, with a median velocity of 320 km/s.   These measures cannot be reliably converted to $v_{\rm rot}$, because we do not know the exact spectral class, we only have a rough limit on the inclination, and because the rotation effect is not large compared to the pressure broadening and thus is hard to pull out in a noisy, variable spectra with partially-filled-in profiles.

Fifth, the companion star must have a fast rotation due to tidal effects forcing it into synchronous rotation.  The rotational synchronization timescale is 390,000 years (Toledano et al. 2007), for an orbital period of 2.04 days, companion of mass 13 $M_{\odot}$, and radius 6.2 $R_{\odot}$.  Given that the age of the FQ Cir binary since the binary orbit got near its current size is likely greatly longer than a quarter-of-a-million years, the companion will have spun-up to synchronize with the orbit.  So the companion is a large star spinning fast.  The equatorial velocity is simply the speed needed to travel the circumference of a 6.2 $R_{\odot}$ star in 2.04 days.  This is $v_{\rm rot}$=150 km/s.

This fast rotation has important implication because this is what makes for gas around the equator to rise above the stellar surface to form the decretion disk.  This is exactly what happens to create the Be stars.  That is, Be stars all have fast rotation, spinning out `decretion disks', extending up to 20--100 $R_{\odot}$ from the stellar surface.  It is the gas in the decretion disk that makes the strong emission lines that define the Be star phenomenon.  Be stars are critically different from FQ Cir, because FQ Cir has the close-in Roche lobe that truncates the size of the disk.  So FQ Cir cannot have a large decretion disk and hence cannot have bright emission lines.  The mass thrown into the decretion disk will ultimately pass through the Roche lobe and fall onto the WD.  This is a new form of acccretion, where the gas flows from a disk around the companion to a disk around the WD.  It is startling to think about a two-disk system, with one disk feeding the other.

If a B-star ever gets into a 2-day orbit with a WD (or NS or BH), then the companion will be spun-up (in a fraction of a million years) into becoming a Be star, will continually eject a decretion disk, and then this gas will speedily go through the Roche lobe, only to go through the accretion disk to fall down onto the compact star.  The decretion disk will be truncated by the small Roche lobe, so the companion can only appear as a {\it weak} Be star, with relatively faint emission lines.

\subsection{Summary of Properties of the FQ Cir binary}

The properties of FQ Cir have many measures, and the critical ones are resolved only at the end.  It is useful to collect all the final and best properties into Table 1.  This will provide the basis for the big questions relating to the nature and evolution of FQ Cir.

\begin{table}
	\centering
	\caption{FQ Cir properties}
	\begin{tabular}{lll} 
		\hline
		 & Property &    Value   \\
		\hline
Nova:	&		&		~		\\
	&	$V_{\rm peak}$	&	10.9$\pm$0.1	~	mag	\\
	&	$B-V$ peak	&	0.68$\pm$0.03	~	mag	\\
	&	$t_2$	&	2	~	days	\\
	&	$t_3$	&	$\le$11	~	days	\\
	&	Light curve	&	S(11)	~		\\
	&	Nova amplitude	&	3.1$\pm$0.1	~	mag	\\
	&	Emission Lines	&	H I, He I	~		\\
	&	Spectral class	&	He/N	~		\\
	&	H$\alpha$ FWHM	&	2600	~	km/s	\\
	&	$M_V$ peak	&	$-$5.8$\pm$0.4	~	mag	\\
Quiescence:	&		&		~		\\
	&	$V_q$	&	14.0$\pm$0.1	~	mag	\\
	&	$B_q-V_q$	&	0.46$\pm$0.03	~	mag	\\
	&	$M_V$	&	$-$2.7$\pm$0.4	~	mag	\\
	&	Absorption lines	&	H I, He I, Na I	~		\\
White dwarf:	&		&		~		\\
	&	$M_{\rm WD}$	&	1.25$\pm$0.10	~	$M_{\odot}$	\\
	&	Initial mass	&	7.6--9.0	~	$M_{\odot}$	\\
	&	Composition	&	ONe	~		\\
Companion:	&		&		~		\\
	&	Classification	&	B1 V(n)(e)	~		\\
	&	Surface temperature	&	22,000$\pm$1000	~	K	\\
	&	Mass $M_{\rm comp}$	&	13.0$^{+0.2}_{-0.5}$	~	$M_{\odot}$	\\
	&	Radius $R_{\rm comp}$	&	6.2$\pm$0.2	~	$R_{\odot}$	\\
	&	Roche lobe $R_{\rm Roche}$	&	1.5$\times$$R_{\rm comp}$	~		\\
	&	Rotation period	&	2.04	~	days	\\
	&	Equatorial velocity	&	150	~	km/s	\\
Binary:	&		&		~		\\
	&	Orbital Period $P$	&	2.041738	~	days	\\
	&	Mass ratio $q$	&	10.4$\pm$0.9	~		\\
	&	Semi-Major axis	&	16.2$\pm$0.1	~	$R_{\odot}$	\\
	&	Extinction $E(B-V)$	&	0.64$\pm$0.05	~	mag	\\
	&	Distance $D$	&	8800$\pm$1500	~	pc	\\
	&	Accretion	&	RLOF from disk	~		\\
	&	Classification	&	HMCV	~		\\		
	
	\hline
	\end{tabular}
\end{table}

\section{HMCV and Disk-RLOF}

\begin{figure}
	\includegraphics[width=1.01\columnwidth]{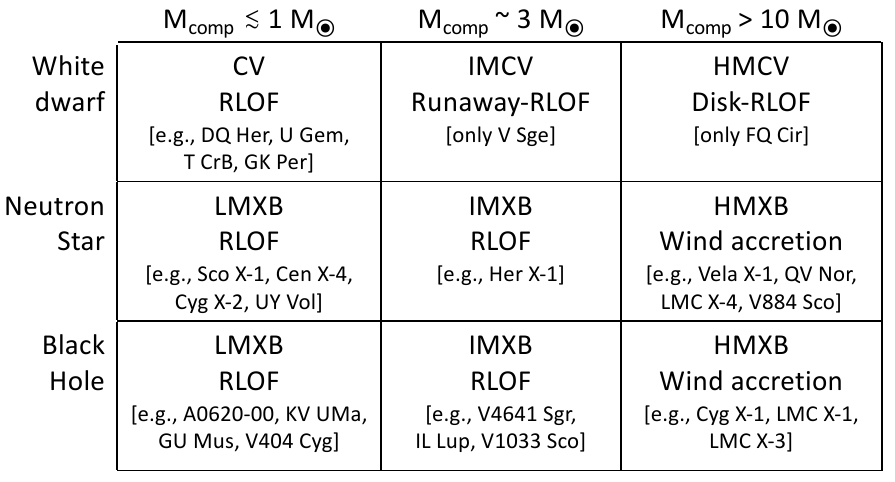}
    \caption{Classification of interacting binaries with collapsed stars.  The classification has been based on the type of collapsed star (WD, NS, and BH) crossed with the mass of the companion star ($\lesssim$1, $\sim$3, and $>$10 $M_{\odot}$).  Interacting binaries with WDs are called cataclysmic variables (CVs), while those with NSs or BHs are called X-ray binaries (XRBs).  Binaries with companions $\lesssim$1 $M_{\odot}$ are called `low mass' (LM), with companions $\sim$3 $M_{\odot}$ are called `intermediate mass' (IM), and with companions $>$10 $M_{\odot}$ are called `high mass' (HM).  Previous to FQ Cir, the box in the upper right for a high mass CV was empty.  Now with FQ Cir, the box has an interesting prototype.  This box needs a name, for which the obvious label is `high mass CV', or HMCV.  }
\end{figure}

FQ Cir is certainly a normal CN occurring in a CV binary with accretion from a 13.0$^{+0.2}_{-0.5}$ $M_{\odot}$ star.  So FQ Cir is a high-mass cataclysmic variable, or an HMCV.  FQ Cir is also a case of a new mode of accretion, where the gas going to the WD is coming by Roche lobe overflow from a decretion disk around the massive star.  This is disk-to-disk accretion.

This discovery of a HMCV fills in a blank on the organization chart for interacting binaries, where a collapsed star is feeding by accretion from some normal donor star.  Figure 9 presents a layout for the three possibilities on the nature of the compact star (WD, NS, or BH) and the three possible ranges for $M_{\rm comp}$ ($\lesssim$1, $\sim$3, and $>$10 $M_{\odot}$).  This makes for nine classes of interacting binaries.  Well known are the CVs with WDs and low-mass companions.  Also well known are the LMXBs, IMXBs, and HMXBs with both NS and BH primary stars.  Few recognize that the utterly unique system V Sge (high mass WD plus a 3.9 $M_{\odot}$ companion with runaway accretion) is the only known `intermediate-mass cataclysmic variable', or IMCV.  Now with FQ Cir, we have a well-measured prototype HMCV.

The nine boxes in Figure 9 are distinct from each other.  The nature of the collapsed star (WD, NS, or BH) has no intermediate states.  The $M_{\rm comp}$ distribution is distinctly tri-modal, with three widely separated modes.  All of the intermediate mass systems that we know about have masses over the narrow range of 2.2$\leq$$M_{\rm comp}$$\leq$3.3 $M_{\odot}$.  We know of no systems of any type within the ranges of 1.0$\lesssim$$M_{\rm comp}$$<$2.2 $M_{\odot}$ or 3.3$<$$M_{\rm comp}$$<$10 $M_{\odot}$.  This observation does not prove such to be impossible, but it must be rare.  So all nine boxes in Figure 9 are distinct, with no overlap or continuum between boxes.  This suggests that there are at least nine separate evolutionary paths that make for interacting binaries with collapsed stars.

We can say that FQ Cir is a WD version of a HMXB.  Or we could say that FQ Cir is a high-mass version of an IMCV (i.e., V Sge).  But neither maxim is useful, because each highlights the similarities of some set of properties while ignoring that several critical properties and mechanisms are greatly different.  Best just to say that FQ Cir is an HMCV.

Amongst the many known HMXBs, a substantial number have B0--B2 main sequence stars, which are Be stars (Walter et al. 2015, Fortin et al. 2023).  The Walter catalog has 10 HMXBs with B0--B2 main-sequence companions, measured orbital periods, and Be star classifications, while the Fortin catalog has 14 more examples with the same constraints.  These all have NS primaries, long orbital periods, and high orbital eccentricities.  These Be HMXBs are greatly different from all CVs due to the NS in the system, with many of them having bright X-ray emission, X-ray pulsars, and cyclotron lines.  Further, the Be HMXBs with known periods have a median near $P$=100 days, while the range is 6.7--262 days.  All of these systems have the companion greatly smaller than their Roche lobe, with the decretion disk truncated by the tidal/resonant interaction with the NS (Fornasini et al. 2023).  All these systems have high orbital eccentricity, and the accretion is episodic as the orbit goes through periastron, so the accretion mode is greatly different from FQ Cir.  So these Be HMXBs differ greatly from FQ Cir in many critical properties and mechanisms.  The Be HMXBs constitute roughly half of the HXMBs, so these systems are relatively common.  Pols et al. (1991) estimate that ``the Be + white dwarf systems should outnumber the Be + neutron star systems by a factor of about 10.''  To the contrary, in our galaxy, there is now only one known HMCV, as compared to 24 Be V HMXBs, so the old estimate is wrong by two orders of magnitude.  We conclude that the evolutionary paths involving high-mass companions are somehow selecting against WDs relative to NSs.  

All the Be HMXBs are are on wide and eccentric orbits while the one HMCV is close to contact.  This might be caused by the NSs receiving natal kicks, while the CVs have common envelopes and RLOF for circular orbits.  We can speculate that this difference in initial binary separation and the corresponding speed of the orbit evolution would make for the HMCVs having a greatly shorter lifetime than Be HMXBs.

Pols et al. (1991) predicted the existence of HMCVs, with the formation of WDs and Be stars through close binary evolution, with the conclusion that such Be+WD systems should be relatively common.  To the surprise of various reviewers (e.g., Rivinius et al. 2013), no such HMCVs have been found in the sky.  Recently, Marino et al. (2025) have claimed to have discovered that an X-ray transient in the Small Magellanic Cloud is a ``Be-white dwarf binary''.  The optical counterpart is an O9--B0e star, conjectured to have some sort of a companion.  Their only evidence for a WD is that ``the X-ray emission resembles the supersoft source phase of typical nova outbursts from an accreting white dwarf (WD) in a binary system.''  But the system certainly did not show any sort of a nova eruption.  Further, their observed X-ray spectrum has critical differences from their ideal WD model (the key `softness' cutoff is 2$\times$ wrong), they worry that the derived temperatures are instrumental artifacts (due to intercalibration between satellites), and they acknowledge that their spectral models are poor matches to the physics of their model.  Further, the mere existence of a soft X-ray spectrum can arise from a wide variety of stars and physical mechanisms (including post-AGB PG 1159 stars, cooling neutron stars, and active galactic nuclei), so the evidence for the WD-conjecture is weak.  Gaudin et al. (2024) make the same claim for the same X-ray transient with the same data, but they add a claimed orbital period of 17.55 days so as to justify the existence of a WD companion.  But the orbital nature of the period is denied by the year-to-year period changes to 17.41 and 17.15 days, such as is impossible for any real binary.  Gaudin et al. try to explain this as a superhump period, but such is denied by the requirement that superhump periods are always $<$3 {\it hours} (Frank et al. 2002).  So there is no orbital period, no WD, and indeed, no evidence for any companion at all.  Six earlier candidate Be+WD systems have been claimed (see Zhu et al. 2023 and references therein), but these are all poorer cases.  All six cases have the only evidence for a WD being ambiguous and problematic X-ray spectra shapes.  None of the six cases have an observed nova eruption to confirm any WD.  None of the cases have a reliable orbital period, so we do not have any useable evidence that the Be star has any companion.  In all, for the seven conjectured Be+WD binaries, we have no useable evidence that the Be star has a WD companion, nor that it has any companion at all.  This is in strong contrast to the case for FQ Cir where the well-measured orbital period proves that the Be star is in a tight orbit around a small or collapsed star, and where the well-observed ordinary classical nova eruption proves that the system has a WD.  So FQ Cir is certainly an HMCV.

Chamoli et al. (2025) published a paper titled ``Challenging Classical Paradigms: Recurrent Nova M31N 2017-01e, a BeWD System in M31?''.  Their abstract starts with ``M31N 2017-01e is the second-fastest recurrent nova known, with a recurrence period of 2.5 yr in the Andromeda galaxy (M31). This system exhibits a unique combination of properties: a low outburst amplitude ($\sim$3 mag), starkly contrasting with known recurrent novae (typically $\ge$6 mag), and a very fast evolution ($t_2$ $\sim$ 5 days). Its position coincides with a bright variable source ($M_V$ $\sim$ $-$4.2, $B-V$ = 0.042) displaying a 14.3 day photometric modulation, which has been suggested as the likely progenitor.''  We judge that the Chamoli et al. paper made good proofs that the star has ordinary thermonuclear eruptions, that the recurrent nova has its quiescent counterpart including the B-star in a 14.3 day binary orbit with the high-mass WD, and that the companion is a high-mass Be star with emission lines that imply a decretion disk.  Further, the Chamoli paper had the convincing insight that the decretion disk must feed the accretion disk around the WD so as to power the eruptions.  They have proven that this rapid recurrent nova is both a HMCV and is a new mode of accretion, from a decretion disk to an accretion disk.

We have independently discovered that FQ Cir is also an example of the two new phenomena;  First, FQ Cir is an example of the long-missing HMCV class.  Second, FQ Cir is an example of a new mode of accretion, involving RLOF operating on a decretion disk of gas, for disk-to-disk transfer.  FQ Cir is bright and relatively close, so the detailed properties and astrophysical situation can be much better measured than for the case of the rapid RN in the Andromeda Galaxy.

\section{Supernova Progenitor?}

FQ Cir is similar to the recurrent nova M31N 2017-01e (with recurrence time of 2.5 years), one of an emerging class of rapid recurrent novae (RRN, Chamoli et al. 2025).  The prototype of the RRNe is M31N 2008-12a (with a one year recurrence time), which is claimed to be ``the leading pre-explosion supernova Type Ia candidate system'' (Darnley et al. 2017).  With these associations, FQ Cir should be considered as a Type Ia supernova (SNIa) progenitor candidate.

The argument for progenitor status does not go beyond the presence of a high-mass WD with a high accretion rate.  This basic argument has been used to tout most classes of CVs as being SNIa progenitors, so repeating it for new classes of CVs (HMCVs and RRNe) carries little conviction.  Well, for FQ Cir, we have no useable evidence to know the accretion rate, but the two RRNe must have high accretion because they have very short recurrence timescales.  More importantly, this basic argument is weak because it ignores the two critical issues of whether the WD ejected mass during a nova eruption ($M_{\rm ejecta}$) is more-or-less than the mass accreted between eruption ($M_{\rm accreted}$), and whether the WD has carbon/oxygen (CO) composition.  The huge problem with checking whether $M_{\rm ejecta}$$>$$M_{\rm accreted}$ is that all the usual and traditional methods for estimating $M_{\rm ejecta}$ have 2--3 orders-of-magnitude real uncertainty\footnote{See Appendix A of Schaefer (2011) and Section 6.1 of Schaefer \& Myers (2025).}.  Importantly, for M31N 2008-12a, we have no measure of $M_{\rm ejecta}$ to any useable accuracy, and the same is true for all the RRNe and FQ Cir.  For the question of the WD composition, a CO WD is required to become a SNIa, with the important alternative being that the WD composition is mainly oxygen/neon (ONe).  ONe WDs are made only from stars with initial masses from 7.0 to 9.0 $M_{\odot}$ (Woosley \& Heger 2015), whereas CO WD are only made from stars with initial masses $<$7.0 $M_{\odot}$.  Observationally, the WD composition can only be determined if neon lines (mainly in the near ultraviolet) appear brightly late in the tail of the nova eruption, as a so-called `neon nova'.  All neon novae must be dredging up and ejecting more mass than was piled on from the accretion, and the neon in startling high abundance can only come from an ONe WD, so for both reasons, neon nova are not ever going to turn into a SNIa.  Unfortunately, the {\it lack} of strong neon emission lines late in the eruption tail does not mean that the WD is of CO composition, because the ONe WDs form with a substantial mantle with little neon (De Ger\'{o}nimo et al. 2019), so the neon would be hidden until many eruptions have stripped away the mantle.  

Neither FQ Cir nor M31N 2017-01e have any spectra useful for seeking neon emission lines.  M31N 2008-12a does not display neon emission in late spectra as seen by {\it HST}, with this being ambiguous as to whether its WD is of CO composition.  So for now, neither FQ Cir nor any of the RRNe have any useful information as to the critical questions (on the ejected mass and the WD composition).

Nevertheless, we can here resolve the question of the WD composition of FQ Cir and M31N 2017-01e.  The analysis centers on the mass budget of the binary, and asking whether the initial mass of the primary star (that will become the WD) can possibly be $<$7.0 $M_{\odot}$.  The beauty of a conservation-of-mass proof is that it does not depend on any history between the initial binary and the current binary.

The original masses of the binary components when they first arrived at the main sequence, $M_{\rm prim}$ and $M_{\rm sec}$, are the primary determinant of the current nature of the WD and the companion.  The initial mass of the primary star must be $M_{\rm prim}$$\leq$9.0 $M_{\odot}$ (Woosley \& Heger 2015) for that star to become a WD.  The initial mass of the secondary star must be smaller than for the primary so that the companion star will still be on the main sequence while the primary has evolved to become a WD.  With detailed consideration of the main sequence (and later) lifetimes\footnote{The current age of the primary in the FQ Cir binary is the time since the coeval formation of the two stars.  The primary's age equals the sum of the main sequence lifetime plus the time to evolve from the main sequence to the time of the WD formation plus the time to in-spiral down to a 2.04 day orbit.  For the primary star, let us take the case of a maximal birth-mass for making a WD that can explode as a supernova of 7 $M_{\odot}$.  The main-sequence lifetime for a 7.0 $M_{\odot}$ star is 38 Myr, while the post-main-sequence duration is 9.6 Myr (Woosley \& Heger 2015).  The time to in-spiral to $P$=2.04 days depends on the $P$ at the end of the planetary nebula phase when the WD is born, which is unlikely to be anywhere near 2.04 days.  The period at the time of WD formation has a wide possible range, but a typical value might be 10 days.  With the universal period change law for HMXBs (Schaefer 2025a), the time from a 10 day orbit to a 2.04 day orbit is 5 Myr.  So the current age in this case is 38+9.6+5 = 52.6 Myr.  For FQ Cir, the companion star (i.e., the secondary star of the original binary) has not yet left the main sequence (even with the late-time addition of many solar masses transferred from the primary), so the main-sequence lifetime for the secondary must be greater than 52.6 Myr.  This requires that $M_{\rm sec}$$<$6.2 $M_{\odot}$ (Woosley \& Heger 2015) for the case of $M_{\rm prim}$=7.0 $M_{\odot}$.  The difference of the original stellar masses must be $>$0.8 $M_{\odot}$ in this extreme case.  For a larger primary mass, the difference in stellar masses will be somewhat larger.}, the primary needs to be at least 0.8 $M_{\odot}$ larger than the secondary, so $M_{\rm prim}$$>$$M_{\rm sec}$$+$0.8 $M_{\odot}$.  A further constraint in the mass budget is to account for the the mass lost by the binary over the eons, $M_{\rm lost}$.  This mass loss comes from ordinary stellar winds, the common envelope phase, the planetary nebula ejection, and the nova ejections.  The ejections are poorly known\footnote{The stellar winds will eject $>$0.21 $M_{\odot}$ (Woosley \& Heger 2015).  Jones et al. (2013) estimates that the primary will lose 0.8 $M_{\odot}$.  For the inevitable common envelope phase (required to get the orbital period down to some small number of days), the primary will transfer mass to the secondary star, but some unknown mass will be flung away from the binary.  The planetary nebula ejection might be anything from 0.2 to 1.0 $M_{\odot}$ or more.  The mass lost to nova eruptions will be roughly the trigger mass of 10$^{-5}$ $M_{\odot}$ with a recurrence time of order 10,000 years for a 1.25 $M_{\odot}$ WD (Yaron et al. 2005) for sometime like 5 Myr, for a total ejected mass of only 0.005 $M_{\odot}$.  For these ejections from the primary star, it is unclear as to what fraction will be captured by the companion star.  A reasonable lower limit for $M_{\rm lost}$ is 0.4 $M_{\odot}$, while an upper limit is poorly known at perhaps 2 $M_{\odot}$.} but must be something like 0.4$\lesssim$$M_{\rm lost}$$\lesssim$2 $M_{\odot}$.   The final constraint on the mass budget is the conservation of mass, with $M_{\rm prim}$$+$$M_{\rm sec}$=$M_{\rm WD}$$+$$M_{\rm comp}$$+$$M_{\rm lost}$.  With the above equations, 2$M_{\rm prim}$$>$$M_{\rm WD}$$+$$M_{\rm comp}$$+$$M_{\rm lost}$$+$0.8 $M_{\odot}$.  For this, we have $M_{\rm WD}$=1.25$\pm$0.10 $M_{\odot}$ and $M_{\rm comp}$=13.0$^{+0.2}_{-0.5}$ $M_{\odot}$.  For a minimal $M_{\rm lost}$, $M_{\rm prim}$$>$7.7$^{+0.1}_{-0.2}$ $M_{\odot}$.  With this mass limit, the primary must have produced an ONe WD.

For advocates of single-degenerate SNIa models wanting HMCVs to be progenitors, the relevant question is ``Can the mass-budget numbers be pushed far enough to allow for the primary star to become a CO WD and possibly a SNIa?''  We can push $M_{\rm comp}$ to 12.5 $M_{\odot}$ and push $M_{\rm WD}$ to 1.15 $M_{\odot}$.  The mass lost to the system is 0.4--2 $M_{\odot}$, so it can be pushed to 0.4 $M_{\odot}$.  The mass difference between the primary and secondary can be pushed from 0.8 $M_{\odot}$ to 0.6 $M_{\odot}$ by presuming that the in-spiral time is zero\footnote{For comparison with footnote 15, we can push to extremes by minimizing the time from the creation of the WD until the time when the orbital period is 2.04 days.  For the case of $M_{\rm prim}$=7.0 $M_{\odot}$, the minimum age of the binary is 38+9.6+0=47.6 Myr.  For the companion to have a main sequence lifetime of $<$47.6 Myr, we must have $M_{\rm sec}$$<$6.4 $M_{\odot}$.  This is a mass difference between the primary and secondary of $>$0.6 $M_{\odot}$.  Changing $M_{\rm prim}$ over the relevant range makes for changes in the mass difference of less than 0.1 $M_{\odot}$.  These minimal mass differences are calculated for isolated stars.  The additional mass added to the companion star late in its life can only make the main sequence lifetime shorter, and hence a mass difference larger than 0.6 or 0.8 $M_{\odot}$.  }.  With these extreme numbers, $M_{\rm prim}$ is 7.32 $M_{\odot}$.  This is to be compared to the largest original mass that can produce a WD with enough carbon to explode as a normal SNIa, with this threshold being near 6.7 $M_{\odot}$\footnote{Woosley \& Heger (2015, figure 2) show that a star with solar abundances originally at 7.0 $M_{\odot}$ will create a WD with only 5\% carbon and have far too little carbon to become a normal SNIa.  They further show that that a 6.5 $M_{\odot}$ star with solar abundances will produce a WD with 32\% carbon content, with a normal SNIa explosion.  So the upper threshold for a primary star to make a WD capable being a SN progenitor is near 6.7 $M_{\odot}$.  This value is not dependent on any binary evolution, because the nuclear burning in the core of the primary does not depend on the secondary, and mass transfer gets significant only at the very end of the primary's post-main-sequence evolution, when it is far too late to matter.  FQ Cir is a relatively young binary, and its spectra are consistent with solar-abundances. }.  So there is no chance for the FQ Cir WD to have a CO composition.  We can also test the {\it unreasonable} extremes, with $M_{\rm lost}$=0 and a zero mass difference, in which case the primary would have its original mass of 6.82 $M_{\odot}$, and still a supernova fate is impossible.  In the end, there is no way to push the numbers so as to allow for FQ Cir to become a SNIa progenitor.  

So the mass-budget requires the original mass of the primary to be $>$7.7$^{+0.1}_{-0.2}$ $M_{\odot}$ and that the FQ Cir WD must be an ONe WD.  There is no possibility that the primary was originally less-massive than 6.7 $M_{\odot}$, so there is no possibility that the WD could have a carbon content adequate to make a normal SNIa.  The final conclusion is that FQ Cir cannot ever become a Type Ia supernova.

The case for M31N 2017-01e (Chamoli et al. 2025) is closely similar to the case for FQ Cir.  One way is to see the model of Marino et al. (2025) for a similar system, with the original mass of the primary being 8 $M_{\odot}$, ending as a 1.23 $M_{\odot}$ ONe WD.  For the case of the M31 RRN specifically, the default scenario is original stellar masses of 9.0 and 8.0 $M_{\odot}$, ending with a 1.37 $M_{\odot}$ ONe WD.  (With a recurrence timescale of 2.5 years, the WD mass must be close to 1.37 $M_{\odot}$.)  The companion is an early B star, more specifically, an early Be star.  The absolute magnitude of the companion is $-$4.2, pointing to a O9 star (Gray \& Corbelly, 2009, Appendix Table B.1).  The SED fits show temperatures from 26,000--31,000 K, which points to a companion from O9.7 to B0.6 (Gray \& Corbally 2009, Appendix Table B.3).  These spectral classifications point to a companion mass from 15--20 $M_{\odot}$ (Cox 2000, Table 15.8).  For the mass-budget analysis, with a minimal 15 $M_{\odot}$ companion, the minimal lost mass of 0.4 $M_{\odot}$, and a minimal primary-secondary mass-difference (to account for the companion still being on the main sequence) of 1.0 $M_{\odot}$, then makes $M_{\rm prim}$=8.9 $M_{\odot}$.  With these pushed values, M31N 2017-01e must now have an ONe WD.  Taking the {\it unreasonable} extremes of $M_{\rm lost}$=0 and the primary-secondary mass-difference to be zero, we see that the original mass of the primary must be $\geq$8.2 $M_{\odot}$.  With this, the mass-budget proves that M31N 2017-01e has an ONe WD, and cannot become a Type Ia supernova.

\section{Evolution of the Binary}

What evolutionary path leads to the observed FQ Cir, and what is its fate?

\subsection{The mass-budget}

The mass-budget analysis (as in Section 4) can tell us about the origin and evolution of FQ Cir.  From the above analysis, the starting mass of the primary star (which becomes the WD) must be between 7.7 and 9.0 $M_{\odot}$.  This tells us that the primary star could only produce an ONe WD.  The original mass of the secondary must be 0.8 $M_{\odot}$ or more smaller than the primary, which is to say that $M_{\rm sec}$ is $<$(6.9--8.2) $M_{\odot}$.  For an extreme $M_{\rm prim}$ and zero-$M_{\rm lost}$, $M_{\rm sec}$ must be $>$5.25 $M_{\odot}$.  The binary evolution has shifted around these original stellar masses, by some mass being lost to the binary ($M_{\rm lost}$) and some mass being transferred from the primary to the secondary.

Let us be specific for the situation with one particular choice of initial stellar masses, 8.5 $M_{\odot}$ for the primary and a 6.5 $M_{\odot}$ companion.  The primary has a main-sequence lifetime of 27 Myr, a post-main-sequence duration of 5 Myr, and an in-spiral time of perhaps 5 Myr.  The current age since formation for this binary is 37 Myr.  The companion has its main-sequence lifetime of 45 Myr.  This lifetime will be modified slightly, because the companion has recently had large masses piled on, and this will shorten the main-sequence evolution somewhat.  For this case, the original total mass in the binary is 8.5$+$6.5=15.0 $M_{\odot}$.  The current total mass in the FQ Cir binary is 13.0$+$1.25=14.25 $M_{\odot}$.  This means that $M_{\rm lost}$ is 15.0$-$14.25=0.75 $M_{\odot}$.  This also means that the secondary star has had 13.0$-$6.5=6.5 $M_{\odot}$ transferred from the primary to the secondary, mostly during the common envelope phase.  With this schematic picture, we have a clear and quantitative scenario for the past evolution of FQ Cir.

Something like this evolution must have taken place to return the current stellar masses.  The exact sequence of events and mechanisms is much more complex than sketched in the last paragraph.  For a comparable case, Marino et al. (2025) present a detailed model that starts with stars of 8 and 6 $M_{\odot}$ in a 3 day orbit, and then proceeds through three giant and common envelope phases and two mass transfer phases to arrive at a 17-day orbital period for a binary consisting of a 1.23 $M_{\odot}$ ONe WD plus a 12 $M_{\odot}$ Be star with a decretion disk.  The mass budget for this model requires that 0.77 $M_{\odot}$ be lost to the binary, and that 6.0 $M_{\odot}$ be transferred from the primary to the companion.  Some such model needs to be specialized for FQ Cir.  While the detailed states and numbers for FQ Cir can be discussed, the overall evolution sketched in the previous paragraph stands as a high-level summary.

\subsection{Evolution from contact to today}

The current binary has a semimajor axis of 16.2$\pm$0.1 $R_{\odot}$ and this is much smaller than the size of the primary star when it was a giant (just before the WD formation).  So there must necessarily have been some sort of a common envelope phase.  This phase is the mechanism by which the $\sim$6 $M_{\odot}$ of mass was transferred to the companion star.  This common envelope is also the mechanism by which the the orbit period was ground down to somewhat larger than 2.04 days.

After the mass transfer stage, the WD and the B-star will have settled down to an orbit longer than  2.04 days in period.  At this time, the binary must be detached, as otherwise the forced exponential rise in accretion would destroy the system within some small number of centuries, while the planetary nebula ejecta was still visible.  Indeed, the binary must remain detached to the present time, as any runaway accretion would be a state lasting only some centuries of time, and we do not see any exponential increase in brightness since 1894.  

After the binary settled down to an orbit of somewhat longer than 2.04 day, within a few hundreds-of-thousands years, the spin of the companion will be forced into synchronous rotation.  With this fast rotation, the companion will have set up oscillations and instabilities that lead to chaotic variability on all timescales, just like for other similar Be fast-rotators.  Further, the fast spin will also amplify magnetic effects that could lead to starspots, flares, and activity cycles.  Just as for all Be stars, this fast rotation of the FQ Cir companion is the cause of the light curve variations on all time scales.

The fast rotation also makes for a decretion disk, just like for the case for all the other Be stars.  This disk will be truncated by the Roche lobe of the binary, thus cutting out most of the emission lines normally seen in Be stars.

Since the end of the last common envelope phase, the orbit has shrunken substantially.  One certain mechanism is the mass transfer.  In general, mass transfer from a star to a less massive star makes the orbital period decrease.  Further, the period change is dominated by some unidentified mechanism for angular momentum loss in high-mass interacting binaries.  (This is certainly not from the venerable mechanism called `magnetic braking', as such has been proven to have either zero or negligibly small effect, see Schaefer 2025a.)  To quantify this, Schaefer (2025a) measured and collected the century-long steady orbital period changes ($\dot{P}$) for 10 HMXBs, plus 67 other XRBs and CVs.  One conclusion was that the angular momentum loss was not a function of the nature of the collapsed star (WD, NS, or BH) in the binary, so these 10 HMXBs should provide close and relevant experience for the period change of FQ Cir.  For the HMXB systems, the $\dot{P}$ is always huge and negative, dominating over other effects.  In an empirical model fit to the HMXBs (equation 17, Schaefer 2025a), the angular momentum loss rate is given as a function of $M_{\rm WD}$, $M_{\rm comp}$, and $P$, with only a very weak dependency on the largely-unknown accretion rate.  With this universal empirical law, the $\dot{P}$ for FQ Cir is estimated to be $-$3.1$\times$10$^{-9}$ in the dimensionless units of s/s.  So the period is decreasing fast.  The evolutionary timescale is $|$$P$/$\dot{P}$$|$ or 910,000 years.  So the original orbit is in-spiraling with a large change over just one-million years.  In the last 2.4 Myr or so, FQ Cir has been in-spiraling from a period of 4 days down to the current 2.04 days.  This evolution brings us to today.  

\subsection{The future evolution}

The current state of the FQ Cir binary is that the period has in-spiraled down to 2.04 days, with the companion star getting close to contacting its Roche lobe.  Over the next 3.4 Myr, the orbit will grind down further until $P$=1.1 days, when the Roche lobe has contracted so much that the companion fills it Roche lobe. 

Repeated nova and helium-flash cycles do not necessarily erode the WD.  Multi-cycle simulations following thousands of outbursts show that white dwarfs can retain most of the transferred mass as the envelope heats and helium burning becomes less degenerate, allowing secular growth toward the Chandrasekhar mass (demonstrated explicitly for CO WDs by Hillman et al. 2016). For CO cores, the expected endpoint is a thermonuclear disruption (i.e., a Type Ia supernova). For ONe cores, current models predict that at similar central densities electron captures trigger accretion-induced collapse (AIC) to a neutron star rather than a normal SNIa.  If one groups AIC and SNIa together as ``catastrophic energetic endpoints'', then an HMCV with an ONe primary remains a plausible progenitor of such an endpoint, contingent on sustained mass retention at high accretion rates. The difference is qualitative (collapse vs thermonuclear disruption) and observational (Ni-56 yield, spectra, and light-curve morphology), not that one outcome is ``energetic'' and the other is not. Hillman et al. (2016) supports the feasibility of the required mass growth; the core composition determines which endpoint occurs.  With FQ Cir having an ONe core, its fate might be the AIC, if its WD mass rises to near the Chandrasekhar limit.

Once the companion star fills its Roche lobe, events start happening fast, with a time scale of centuries.  The situation becomes complex, and there is substantial uncertainty in the results from the known mechanisms.  Certainly for the first few centuries after contact, the well-known dynamical instability will drive an exponentially runaway accretion.  The e-folding time for this runaway is likely to be something on the order of a century.  The WD might keep up with the accretion by having frequent nova eruptions.  When the accretion rate rises above 2$\times$10$^{-7}$ $M_{\odot}$ yr$^{-1}$, steady hydrogen burning on the WD surface commences (Shen \& Bildsten 2009), making for a luminous hot X-ray source.  When the accretion rate rises above 6$\times$10$^{-7}$ $M_{\odot}$ yr$^{-1}$, the accreted gas will expand to form an envelope with red giant proportions (Shen \& Bildsten 2009).  This envelope will hide the nuclear burning in the interior.  This envelope will be much larger than the binary orbit, so we will have a common-envelope situation with the companion.  This will only increase the rate of mass falling onto the WD even higher.  Sooner or later, the mass in the WD core will rise to near the Chandrasekhar limit.  The WD has an ONe composition, so there can be no supernova explosion.  Rather, the core will undergo an accretion-induced collapse, quietly forming a NS in the middle of what is left of the binary and envelope.  After the collapse to a NS, we might be left with an HMXB.  That is, the accretion would still be on-going, and the NS would have a high-mass companion.  To name a particular HMXB analog, we can point to the X-ray pulsar Cen X-3, with its 2.087 day period and a companion star (Krzemi\'{n}ski's Star) that is 24 $M_{\odot}$.  This possibility is the case where an HMCV is a progenitor for an HMXB.

Past the formation of an HMXB with the accretion induced collapse, the further evolution of FQ Cir is unclear.  The NS might turn into an X-ray pulsar.  Judging from the other HMXBs with measured $\dot{P}$, the FQ Cir binary would be fast in-spiraling to keep a high accretion rate.  If the accretion continues, perhaps the NS will collect enough mass to collapse into a BH.  Perhaps we will end up with an HMXB with a black hole, still accreting mass at a high rate.  Perhaps the future evolution of such a BH HMXB is for most of the companion mass to be swallowed by the black hole, until the companion is evaporated away or until the accretion turns off.  The speculation in this paragraph is schematic, other channels might be better, and all of this is incredibly far in the future.

\section{Nature of the System}

\begin{figure*}
	\includegraphics[width=2.11\columnwidth]{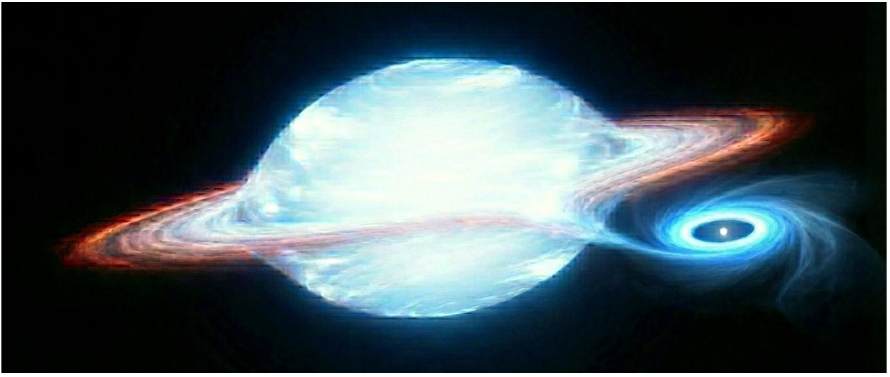}
    \caption{Schematic illustration of the unique FQ Cir binary.  This figure was made by B. Tremblay (Vatican Observatory), and is a reasonable close-up view of the binary.  We see the hot B1e star filling just over half of its Roche lobe, while its fast rotation makes for a rather oblate shape.  We see the relatively narrow decretion disk being ejected around the equatorial regions of the hot star.  This decretion disk spreads out within the Roche lobe, where gas near the L1 point will fall down onto the white dwarf.  The falling gas will form an accretion stream, then form an ordinary accretion disk around the WD.  The gas fallen onto the WD will accumulate, eventually reaching the hydrogen ignition point, and erupt as a normal nova eruption.  This picture illustrates the two unique properties of FQ Cir, first that the companion is a high mass star making this a High-Mass CV (HMCV), and second that we have the entirely new accretion mode where one disk feeds a second disk.  }
\end{figure*}

The nature of FQ Cir has become well-measured and clear.  Figure 10 copies an illustration showing the two stars and the two disks, with this for helping readers to visualize the unique and unfamiliar layout of FQ Cir.  The WD is massive, near 1.25 $M_{\odot}$, so the nova eruption is of the He/N spectral class with a fast decline ($t_2$=2 days and $t_3$$\le$11 days) and broad Balmer lines (FWHM=2600 km/s).  The WD must have ONe composition, and can never become a Type Ia supernova.  The quiescent nova counterpart is certainly the 14th-mag B-star, where the luminous companion is what makes for the small eruption amplitude of 3.1 mag.  The orbital period is 2.04 days, and the binary motion has spun up the companion star to a fast rotation rate (with equatorial velocity of 150 km/s) from spin/orbit synchronization.  With this fast rotation, the companion star ejects gas in an equatorial decretion disk, just as in the common Be stars (with their full decretion disks typically out to 20--40 $R_{\odot}$ making for bright emission lines).  The companion is a B1 V(n)(e) star (13.0 $M_{\odot}$, 22,000$\pm$1000 K, 6.2 $R_{\odot}$).  The companion does not fill up its Roche lobe, with the Roche lobe being 1.5$\times$ the stellar radius.  The ejected decretion disk is truncated by the Roche lobe, so the system does not show bright emission lines, rather we only see weak and intermittent emission from the small decretion disk.  All the gas that goes into the decretion disk will ultimately cross outside the Roche lobe and fall into the accretion disk around the WD, then fall onto the WD to power the ordinary nova eruption.  Like Be stars, the underlying fast-rotating Be-star light curve undergoes trends, flares, and flickering on all time scales, at up to the half-magnitude level.  This fast rotation makes for the somewhat-broadened Balmer lines, which are sometimes partly filled in with line emission from the small truncated decretion disk.  In summary, FQ Cir had an ordinary nova eruption from a relatively high-mass ONe WD, however the binary has two new and startling features, being both a high-mass cataclysmic variable, and a new form of accretion where Roche lobe overflow from the decretion disk around the Be star companion feeds the normal accretion disk around the white dwarf.

\section{Further Study}

What are the tasks for future study?  {\bf A.~}The most obvious need is to get a full radial velocity curve.  This will help pin down the exact masses and inclination.  This study would also provide some monitoring of the variations and velocities of the emission that partly fills the Balmer lines.  {\bf B.~}It would be good to obtain a spectrum with high signal-to-noise (say, $>$100), with good spectral resolution (say, $R$$>$20,000), and with coverage over the blue (say, 3800--4600~\AA).  This will allow an exacting measure of the temperature, luminosity class, and perhaps even $v \sin i$.  This might pull out emission lines from the accretion disk around the WD, giving some measure of the accretion rate.  The same observing run can also cover the H$\alpha$ profile to measure velocities of features in the emission.  {\bf C.~}Another type of new observation that might be useful is X-ray spectroscopy, with the hope of getting information on the accretion low down near the WD.  This task might be difficult, because the target is at a distance of 8800 pc (like to the Galactic center) so the system might be too faint to detect unless the accretion rate is high.  {\bf D.~}Now that we know what to look for, further HMCV systems might be identified from archival records of suspected `dwarf nova', seen only once (or twice!), that have high-luminosity blue counterparts.  {\bf E.~}With optimal filtering of nightly transients, new HMCVs can be discovered in near-real-time with the Vera Rubin Observatory sky-survey.  By itself, Rubin cannot provide adequate follow-up for these short-duration transients, so the transient-alert brokers must somehow highlight the new HMCV-nova as being extremely-rare and worthy of massive follow-up.  {\bf F.~}FQ Cir represents an entirely new path of binary evolution, and a full detailed model specific to FQ Cir is needed.  This study should include addressing why there are so few HMCVs.  {\bf G.~}And finally, some sort of a theory study should be made of this new mode of accretion.

\begin{acknowledgments}

We appreciate the illustration of the FQ Cir binary (Figure 10) contributed by Bob Tremblay (Vatican Observatory, Warren Astronomical Society).  SJM was supported by the Australian Research Council through Future Fellowship FT210100485.  LJT acknowledges support from the South African National Research Foundation.  Some of the data presented in this paper were obtained with the Southern African Large Telescope under the program 2025-2-DDT-005 (PI: Shara).  We are grateful to Alexei Kniazev for helping to correct an error in the initial SALT/HRS data reduction. 

\end{acknowledgments}

%



{}


\end{document}